\pdfoutput=1
\documentclass[11pt,a4paper]{article}
\usepackage[left=2.5cm,right=2.5cm,top=2.8cm,bottom=2.8cm]{geometry}
\usepackage{amsmath,amssymb,amsthm,amscd}
\usepackage{mathtools}
\usepackage{jheppub}
\usepackage{xcolor}
\usepackage{colortbl}
\usepackage{bbold}
\usepackage{url}
\usepackage{empheq}
\usepackage{graphicx}
\usepackage{nicematrix}
\usepackage{float}
\usepackage{fontawesome}
\usepackage{tikz}
\usetikzlibrary{calc}
\usetikzlibrary{patterns}
\usetikzlibrary{shapes.misc}
\usetikzlibrary{
    decorations.pathmorphing,
    decorations.pathreplacing,
    decorations.shapes,
    decorations.markings,
}
\usetikzlibrary{fit, shapes, tikzmark}

\tikzset{ shorten <>/.style={ shorten >=#1, shorten <=#1 } }
\makeatletter
\tikzset{
    dot diameter/.store in=\dot@diameter,
    dot diameter=3pt,
    dot spacing/.store in=\dot@spacing,
    dot spacing=10pt,
    dots/.style={
        line width=\dot@diameter,
        line cap=round,
        dash pattern=on 0pt off \dot@spacing
    },
    mid/.style={
        postaction={
            decorate,
            decoration={
                markings,
                mark=at position .9 with
                {\arrow{stealth}}
            }
        }
    },
}
\makeatother

\makeatletter
\@ifundefined{varclubsuit}{%
  \def\re@DeclareMathSymbol#1#2#3#4{%
    \let#1=\undefined
    \DeclareMathSymbol{#1}{#2}{#3}{#4}}%
  \DeclareSymbolFont{symbolsC}{U}{txsyc}{m}{n}%
  \SetSymbolFont{symbolsC}{bold}{U}{txsyc}{bx}{n}%
  \re@DeclareMathSymbol{\varclubsuit}{\mathord}{symbolsC}{112}%
  \re@DeclareMathSymbol{\vardiamondsuit}{\mathord}{symbolsC}{113}%
  \re@DeclareMathSymbol{\varheartsuit}{\mathord}{symbolsC}{114}%
  \re@DeclareMathSymbol{\varspadesuit}{\mathord}{symbolsC}{115}%
}{}

\makeatother

\usepackage{orcidlink}
\usepackage{lmodern}
\usepackage[T1]{fontenc}
\usepackage{microtype}
\usepackage{bbm}
\usepackage{faktor}

\newcommand{\brk}[1]{(#1)}
\newcommand{\lrbrk}[1]{\left(#1\right)}
\newcommand{\bigbrk}[1]{\bigl(#1\bigr)}
\newcommand{\Bigbrk}[1]{\Bigl(#1\Bigr)}

\newcommand{\sbrk}[1]{[#1]}

\newcommand{\brc}[1]{\{#1\}}

\newcommand{\abs}[1]{|#1|}
\newcommand{\vev}[1]{\langle #1\rangle}
\newcommand{\bigvev}[1]{\big\langle #1\big\rangle}
\newcommand{\Bigvev}[1]{\Big\langle #1\Big\rangle}

\newcommand{\bigfloor}[1]{\big\lfloor #1 \big\rfloor}
\newcommand{\Bigfloor}[1]{\Big\lfloor #1 \Big\rfloor}

\newcommand{\supbrk}[1]{^{\brk{#1}}}
\newcommand{\brkbf}[1]{{\brk{\mathbf{#1}}}}
\newcommand{\supbbf}[1]{^{\brk{\mathbf{#1}}}}

\newcommand{\bra}[1]{\langle #1 |}
\newcommand{\bigbra}[1]{\big\langle #1 \big|}
\newcommand{\Bigbra}[1]{\Big\langle #1 \Big|}

\newcommand{\ket}[1]{| #1 \rangle}
\newcommand{\bigket}[1]{\big| #1 \big\rangle}
\newcommand{\Bigket}[1]{\Big| #1 \Big\rangle}

\DeclarePairedDelimiterX\braket[2]{\langle}{\rangle}{#1 \>\delimsize\vert\> #2}

\newcommand{\dd}{\mathrm{d}}
\newcommand{\dlog}{\dd \log}
\newcommand{\defas}{:=}

\newcommand{\Integers}{\mathbb{Z}}

\newcommand{\Complex}{\mathbb{C}}
\newcommand{\Field}{\mathbb{K}}

\newcommand{\Poles}{\mathcal{P}}

\newcommand{\projPhi}{\varphi}
\newcommand{\Baikov}{\mathcal{B}}
\newcommand{\Res}{{\rm Res}}
\newcommand{\Cmat}{{\bf C}}

\newcommand{\rmH}{\mathrm{H}}
\newcommand{\Dim}{\mathrm{dim}}

\makeatletter
    \newcommand*\bigcdot{\mathpalette\bigcdot@{1}}
    \newcommand*\smtimes{\mathpalette\smtimes@{.7}}
    \newcommand*\bigcdot@[2]{\mathbin{\vcenter{\hbox{\scalebox{#2}{$\m@th#1\bullet$}}}}}
    \newcommand*\smtimes@[2]{\mathbin{\vcenter{\hbox{\scalebox{#2}{$\m@th#1\times$}}}}}
\makeatother

\newcommand{\fun}[2]{\code{#1[}\,#2\,\code{]}}

\newcommand{\mId}{\mathbb{1}}
\newcommand{\mzero}{\gr{0}}
\newcommand{\mzeroRed}{\color{red!25}{0}}
\newcommand{\mzeroRedd}{\color{red!10}{0}}
\newcommand{\Span}{\operatorname{Span}}
\newcommand{\Deg}{\operatorname{deg}}
\newcommand{\tr}{^T}
\newcommand{\cmatQ}[1]{Q_{#1}}

\newcommand{\ctensor}[1]{\cT_{#1}}

\newcommand{\cL}{\mathcal{L}}

\newcommand{\cQ}{\mathcal{Q}}

\newcommand{\cT}{\mathcal{T}}

\newcommand{\cX}{\mathcal{X}}

\newcommand{\soft}[1]{\textsc{#1}}

\newcommand{\filename}[1]{\texttt{#1}}
\newcommand{\code}[1]{\texttt{#1}}
\newcommand{\wavy}[1]{
    \begin{tikzpicture}[baseline=-0.5ex]
        \coordinate (origin) at (0, 0);
        \draw[->,
            decorate, decoration={
                snake,
                segment length=.2cm,
                amplitude=.03cm
            }
        ] (origin) -- +(#1, 0);
    \end{tikzpicture}
}

\newcommand{\namedref}[2]{\hyperref[#2]{#1~\ref*{#2}}}
\newcommand{\secref}[1]{\namedref{Section}{#1}}

\newcommand{\appref}[1]{\namedref{Appendix}{#1}}

\newcommand{\figref}[1]{\namedref{Figure}{#1}}

\makeatletter
\def\mr@ignsp#1 {\ifx\:#1\@empty\else #1\expandafter\mr@ignsp\fi}%
\newcommand{\multiref}[1]{\begingroup
\xdef\mr@no@sparg{\expandafter\mr@ignsp#1 \: }%
\def\mr@comma{}%
\@for\mr@refs:=\mr@no@sparg\do{\mr@comma\def\mr@comma{,\,}\ref{\mr@refs}}%
\endgroup}
\makeatother
\renewcommand{\eqref}[1]{(\multiref{#1})}

\newcommand{\be}{\begin{equation}}
\newcommand{\ee}{\end{equation}}
\newcommand{\bea}{\begin{eqnarray}}
\newcommand{\eea}{\end{eqnarray}}
\newcommand{\bei}{\begin{itemize}}
\newcommand{\eei}{\end{itemize}}

\makeatletter
\newlength{\apb@width}
\newcommand{\autoparbox}[2][c]{\settowidth{\apb@width}{#2}\parbox[#1]{\apb@width}{#2}}
\newcommand{\includegraphicsbox}[2][]{\autoparbox{\includegraphics[#1]{#2}}}
\makeatother

\definecolor{green1}{HTML}{244819}
\definecolor{cyan1}{HTML}{37cdaa}
\definecolor{blue1}{HTML}{5d7ac4}
\definecolor{red1}{HTML}{d0482a}
\definecolor{purple1}{HTML}{845ea8}
\definecolor{orange1}{HTML}{e07229}
\definecolor{yellow1}{HTML}{edcb52}
\definecolor{red}{HTML}{921818}
\definecolor{purple}{HTML}{53047A}
\definecolor{yellow}{HTML}{f4e097}
\definecolor{gr}{gray}{0.7}
\definecolor{gr1}{gray}{0.7}
\newcommand{\gr}[1]{{\color{gr}#1}}

\usepackage{etoolbox}
\makeatletter
    \newcommand\newsubsupcommand[4]{\newcommand#1{#2\sc@subp{#3}{#4}}}
    \def\sc@subp#1#2{%
        \let\sc@subflag\undefinded%
        \let\sc@supflag\undefinded%
        \def\sc@thesub{#1}%
        \def\sc@thesup{#2}%
        \sc@proc%
    }%
    \def\sc@proc{%
        \@ifnextchar{_}{\def\sc@subflag{}\sc@mergesubs}{%
            \@ifnextchar{^}{\def\sc@supflag{}\sc@mergesups}{
                \ifdef{\sc@subflag}{}{_{\sc@thesub}}%
                \ifdef{\sc@supflag}{}{^{\sc@thesup}}%
            }%
        }%
    }%
    \def\sc@mergesubs#1#2{_{\sc@thesub#2}\sc@proc}%
    \def\sc@mergesups#1#2{^{\sc@thesup#2}\sc@proc}%
\makeatother

\newsubsupcommand{\phiL}{\varphi}{}{}
\newsubsupcommand{\hatPhiL}{\widehat{\varphi}}{}{}
\newsubsupcommand{\tphiL}{\tilde{\varphi}}{L}{}
\newsubsupcommand{\phiR}{\varphi}{}{\vee}
\newsubsupcommand{\hatPhiR}{\widehat{\varphi}}{}{\vee}
\newsubsupcommand{\psiL}{\psi}{}{}
\newsubsupcommand{\psiR}{\psi}{}{\vee}
\newsubsupcommand{\vphiL}{\vec{\varphi}}{}{}
\newsubsupcommand{\vpsiL}{\vec{\psi}}{}{}
\newsubsupcommand{\vphiR}{\vec{\varphi}}{R}{}

\newcommand{\muL}{\mu}
\newcommand{\muR}{\mu^\vee}

\def\seva#1{}
\def\pierpaolo#1{}
\def\giacomo#1{}

\title{Intersection Numbers from Companion Tensor Algebra}

\author[a,b,\orcidlink{0009-0004-4788-738X}]{Giacomo Brunello,}
\author[c,\orcidlink{0000-0001-7067-0315}]{Vsevolod Chestnov,}
\author[a,\orcidlink{0000-0001-9711-7798}]{Pierpaolo Mastrolia}

\newcommand{\padova}{
    Dipartimento di Fisica e Astronomia, Universit\`a degli Studi di Padova
    e INFN, Sezione di Padova,
    Via Marzolo 8, I-35131 Padova, Italy.
}

\newcommand{\bologna}{
    Dipartimento di Fisica e Astronomia, Universit\`a di Bologna
    e INFN, Sezione di Bologna,
    via Irnerio 46, I-40126 Bologna, Italy.
}

\affiliation[a]{\padova}
\affiliation[b]{Institut de Physique Théorique, CEA, CNRS, Université Paris-Saclay, F–91191 Gif-sur-Yvette cedex, France}
\affiliation[c]{\bologna}

\emailAdd{giacomo.brunello@phd.unipd.it}
\emailAdd{vsevolod.chestnov@unibo.it}
\emailAdd{pierpaolo.mastrolia@unipd.it}

\abstract{
    Twisted period integrals are ubiquitous in theoretical physics and
    mathematics, where they inhabit a finite-dimensional vector space governed
    by an inner product known as the intersection number.
    In this work, we uncover the associated tensor structures of intersection
    numbers and integrate them with the fibration method to develop a novel
    evaluation scheme.
    Companion matrices allow us to cast the computation of the intersection
    numbers in terms of a matrix operator calculus within the ambient tensor
    space.
    For illustrative purposes, our algorithm has been successfully applied to
    the numerical decomposition of a sample of two-loop integrals,
    coming from planar five-point massless functions, representing a
    significant advancement for the direct projection of Feynman integrals to
    master integrals via intersection numbers.
}

\begin{document}

\maketitle

\section{Introduction}
Understanding the properties of \textit{twisted period integrals} is of fundamental importance
in theoretical physics and mathematics, as they appear in a plethora of
applications, ranging from Correlation Functions in Classical and Quantum Field
Theory, Feynman Integral and Scattering Amplitudes computation, Cosmological
Wavefunctions of the Universe, String Theory, and Gravitational Waves Physics,
as well as in the formal studies of special functions like the Aomoto-Gau\ss{}
Hypergeometric functions, Euler-Mellin integrals, and
Gelfand-Kapranov-Zelevinsky systems, and their applications in the realm of
differential geometry, combinatorics, and statistics.

Defined as integrals of a differential $n$-form weighted by a regulated
multivalued function, known as the \textit{twist}, that vanishes at the boundary
of the integration domain, twisted period integrals admit a finite-dimensional vector space
structure,
whose characteristics can be explored using the framework of
intersection theory for twisted cohomology~\cite{%
    matsumoto1994, matsumoto1998, OST2003, doi:10.1142/S0129167X13500948,%
    goto2015, goto2015b, Yoshiaki-GOTO2015203, Mizera:2017rqa,%
    matsubaraheo2019algorithm, Ohara98intersectionnumbers,%
    https://doi.org/10.48550/arxiv.2006.07848,%
    https://doi.org/10.48550/arxiv.2008.03176,%
    https://doi.org/10.48550/arxiv.2104.12584%
}.
Differential $n$-forms live in a vector space defined as the quotient space of the closed
forms {\it modulo} the exact forms, known as twisted de Rham cohomology group
\cite{Mastrolia:2018uzb,Frellesvig:2019kgj,Frellesvig:2019uqt,Frellesvig:2020qot}. The
dimension, as well as the algebraic and analytic properties of this group, depend on the
geometry of the twist, namely on its zeroes and its critical points.
Once a basis of forms spanning this space has been introduced, it is possible
to derive linear and quadratic relations among the elements, as well as
differential and difference equations.
The elementary quantities that define such relations, are given by
\textit{intersection numbers} of meromorphic differential  $n$-forms, which act
as an inner product in the cohomology space.
Developing optimal algorithms for calculating intersection numbers for
meromorphic twisted $n$-forms remains a challenge of significant interest to
both mathematicians and physicists, and different methods have been developed,
leveraging the twisted version of the Stokes' theorem~\cite{cho1995}.
The most efficient approach to evaluate $n$-form intersection numbers nowadays,
as detailed in~\cite{Mizera:2019gea,Frellesvig:2019uqt,Frellesvig:2020qot}, is
based on the concept of \textit{fibration}~\cite{Ohara98intersectionnumbers},
an approach that amounts to recursive evaluation of intersection numbers for
vector-valued 1-forms, for which a systematic algorithm is
known~\cite{Mastrolia:2018uzb}.
Improving the evaluation of 1-form intersection numbers is then of fundamental
importance in order to make the fibration algorithm more efficient.
One simplification arises from exploiting the invariance of the representative of
the cohomology classes and by avoiding the use of algebraic extensions~\cite{Weinzierl:2020xyy}.
Singularities not regulated by the twist were originally addressed by adding
extra analytic regulators \cite{Frellesvig:2019uqt,Frellesvig:2020qot}, which
made computations more cumbersone.
Remarkably, such singularities are naturally incorporated in the framework of
\textit{relative twisted cohomology}~\cite{matsumoto2018relative}, and are managed using multivariate Leray
coboundaries, eliminating the need for additional
parameters~\cite{Caron-Huot:2021iev,Caron-Huot:2021xqj}.
A noteworthy enhancement in this evaluation process, which avoids the
appearance of algebraic extension, has been introduced in
\cite{Fontana:2023amt} and applied in~\cite{Brunello:2023rpq}, exploiting
techniques from computational algebraic geometry such as \textit{polynomial
ideals} and the global residue theorem, and employing the technology of the finite fields
reconstruction of rational functions~\cite{Peraro:2016wsq,Peraro:2019svx}.
As an alternative to the fibration technique, a direct computation method
has been implemented in the logarithmic case
\cite{matsumoto1998,Mizera:2017rqa,Mizera:2019vvs,Matsubara-Heo:2021dtm}, while
for generic meromorphic $n$-forms only partial results are known in literature
\cite{cho1995, matsumoto1998,Chestnov:2022xsy}.

It has been recently proposed another technique~\cite{Chestnov:2022alh}, based on the solution of the \textit{secondary
equation} built from the \textit{Pfaffian system} of differential equations for
the generators of the cohomology group
\cite{Matsubara-Heo-Takayama-2020b,matsubaraheo2019algorithm}, obtained via an efficient algorithm for construction of such systems by means of the
Macaulay matrix \brk{see also~\cite{Henn:2023tbo} for another application}.
In mathematics, intersection theory has been widely used to study the vector space
structure of Aomoto-Gelfand and Gau\ss{} hypergeometric integrals, as well as
Gelfand-Kapranov-Zelevinsky hypergeometric systems~\cite{GKZ-1989,GKZ-Euler-1990,Chestnov:2022alh}.
Our main focus in developing
intersection theory is to apply it to Feynman integrals~\cite{Mastrolia:2018uzb}, which are
among the key elements in perturbative Quantum Field Theory \brk{pQFT}, and can be seen as
singular limit of the former mathematical structures~\cite{Chestnov:2023kww}
(see also~\cite{Agostini:2022cgv,Matsubara-Heo:2023ylc}).
After a systematic change to a parametric representation
\cite{Baikov:1996iu,Lee:2013hzt,Frellesvig:2017aai,Mastrolia:2018uzb}, the
twisted period integral structure becomes evident, and $n$-forms of
dimensionally regulated Feynman integrals become elements of a relative
cohomology group, with relative surfaces corresponding to the physical propagators.
In this framework, linear relations among differential $n$-forms are known as
\textit{integration-by-parts} (IBP) identities
\cite{Tkachov:1981wb,Chetyrkin:1981qh}, in which integrals are decomposed in
terms of a basis of master integrals (MIs).
These integrals obey differential equations, whose canonical
form~\cite{Henn:2013pwa, Chen:2020uyk, Chen:2022lzr, Chen:2023kgw,Giroux:2022wav} (in
presence of generalised polylogarithms as well as of elliptic functions), properties~\cite{Duhr:2023bku}, and
their symmetries~\cite{Pogel:2024sdi,Duhr:2024xsy} can be investigated in terms of
intersection numbers.
Moreover, the symbol letters, which are related to the singularities appearing in the DEs,
have been studied using intersection theory alongside a novel understanding
of the recursive structure of the Baikov representation
\cite{Jiang:2023qnl,Jiang:2023oyq,Jiang:2024eaj}.
Besides that, quadratic relations among FIs are known as twisted Riemann
bilinear relations, and they can also be investigated by means of intersection
numbers~\cite{Mastrolia:2022tww,Matsumoto:2022Va,Duhr:2024rxe}.
In the previous work~\cite{Brunello:2023rpq}, these techniques were applied to the decomposition of two-loop four-point Feynman Integrals,
planar and non-planar, massless and with one external mass, requiring the
evaluation of 9-form intersection numbers.
Beyond FIs, twisted cohomology found many new applications in physics: it can
be applied to Correlation Functions in quantum mechanics
\cite{Cacciatori:2022mbi}, and lattice gauge theory
\cite{Weinzierl:2020gda,Weinzierl:2020nhw,Gasparotto:2022mmp,Gasparotto:2023roh}.
Recent studied showed the applicability also to Fourier transform
\cite{Brunello:2023fef}, seen as confluent hypergeometric function
\cite{Matsumoto1998-2,majima2000} such as those appearing in
gravitational waves waveforms computations~\cite{Brunello:2024ibk}\footnote{
    See also~\cite{Frellesvig:2024swj} for another application to post-Minkowskian integrals
}.
Cosmological wavefunctions of the Universe~\cite{Arkani-Hamed:2017fdk}, that
for years have been computed using standard pQFT techniques, are now being
understood to have a twisted period integral
structure~\cite{De:2023xue,Arkani-Hamed:2023kig,Arkani-Hamed:2023bsv,Benincasa:2024ptf},
where the intersection theory can be applied.

In this work, we propose an improved framework to evaluate $n$-form intersection
numbers that reveals the underlying tensor structures.

Evaluation of 1-form intersection numbers requires
local holomorphic solutions of differential equations near
the singularities of the connection. Usually, these solutions
are found by constructing ans\"atze at each singularity.
However, as shown in~\cite{Fontana:2023amt}, this process can be bypassed using
polynomial division method and the global residue theorem. By introducing an
appropriate polynomial ideal~\cite{Brunello:2023rpq}, whose vanishing set
encompasses the singularities of the connection, and choosing an element of the
quotient ring as the new unique ansatz, we simplify the computation.

We advance this idea by treating the ansatz as an element of a tensor
space composed of three factors: the vector-valued cohomology of the
fibration layer, the quotient ring, and the space of Laurent expansions around
the vanishing set of the polynomial ideal.
Adopting this approach, we reformulate the computation of intersection
numbers in the language of matrix calculus. 
Companion matrices, which naturally encode the polynomial division operation in
terms of matrix multiplication, are central to this reformulation.
Additionally, the Weyl algebra of differential operators gets represented by
infinite-dimensional operators acting on the space of
Laurent expansions.

By combining these two representations, the differential equation
is reinterpreted through tensor product companion matrices with four
indices acting on the ansatz, while the global residue is translated into a covector
on the tensor space.
This approach not only illuminates the mathematical properties and patterns of
intersection numbers, but also greatly enhances computational performance.
The primary advantage of the proposed algorithm lies in its reliance on
algebraic and matrix operations, which can be efficiently coupled with finite
field reconstruction techniques.

To demonstrate the effectiveness of our method, we apply it to the numerical decomposition
of two-loop five-point functions, which require the computation of $11$-form
intersection numbers. This application marks a significant milestone in the full
decomposition of two-loop Feynman integrals using projection through
intersection numbers.

The work is organised as follows:
in~\secref{sec:review} we give an overview of intersection numbers for relative
twisted de Rham cohomology, and of the twisted period integral structure of
Feynman integrals.
In \secref{sec:ctensor} we introduce the tensor structure of intersection
numbers and the companion matrix representation.
In~\secref{sec:pentabox} we show the effectiveness of our novel method by
applying it to the reduction of massless two-loop five-point functions.
We provide concluding remarks in \secref{sec:outlooks}.

The manuscript contains three appendices:
in \appref{app:polydiv} we show that polynomial division operation can be read
as a change of coordinates.
In \appref{app:example} we show a pedagogical example in which we carry out a
step-by-step evaluation of intersection numbers in a univariate example,
and in \appref{app:code} we give details on the computer implementation of our
framework.

For our research, the following software has been used:
\soft{LiteRed}~\cite{Lee:2012cn, Lee:2013mka},
\soft{FiniteFlow}~\cite{Peraro:2019svx},
\soft{Fermat}~\cite{Fermat},
\soft{Fermatica}~\cite{Fermatica},
\soft{Mathematica},
\soft{Singular}~\cite{DGPS} and its \soft{Mathematica} interface {\sc Singular.m}~\cite{SingularInterface}, and
\soft{Jaxodraw}~\cite{Binosi:2008ig, Axodraw:1994}.

\section{Relative Twisted Cohomology and Intersection Theory}
\label{sec:review}
In this section we review in a pragmatic way the main concepts of twisted
cohomology and intersection theory, focusing on the computational aspects of
$1$- and $n$-form intersection numbers, in the
framework~\cite{Brunello:2023rpq} combining the \textit{global residue
theorem}~\cite{Weinzierl:2020xyy},
the \textit{polynomial division} method~\cite{Fontana:2023amt},
and \textit{relative twisted cohomology}~\cite{Caron-Huot:2021iev,Caron-Huot:2021xqj}.

\subsection{Twisted period integrals}
Objects of our study are \textit{twisted period integrals} and their duals,
defined as bilinear products between differential $n$-forms: $\phiL =
{\widehat \phiL(z)} \, \dd^n z \equiv {\widehat \phiL}(z) \, \dd z_1
\wedge \ldots \wedge \dd z_n$ \brk{similarly for the dual form $\phiL^\vee$}, and
$n$-dimensional integration domains $\mathcal{C}$ and $ \mathcal{C}^\vee$:
\begin{align}
    I = \int_{\mathcal{C}} \! u \ \phiL \defas \langle \phiL | \mathcal{C} ]
    \>, \qquad
    I^{\vee} = \int_{\mathcal{C}^\vee} \! u^{-1} \ \phiR \defas [ \mathcal{C}^\vee | \phiR \rangle
    \>, \qquad
    u = \prod_i B_i^{\gamma_i}
    \>,
    \label{eq:integraldef}
\end{align}
where $u$ is a multivalued function called the \textit{twist}, with polynomial
factors $B_i = B_i(z)$ and generic exponents $\gamma_i\>$.

The integration domains are defined in such a way that the twist vanishes on
their boundaries: $\prod_i B_i(\partial \mathcal{C})=\prod_i B_i(\partial
\mathcal{C}^{\vee})=0$, ensuring that the integrals~\eqref{eq:integraldef} obey
integration-by-parts (IBPs) identities:
\begin{align}
\label{eq:ibpi}
    \bigbra{\nabla_\omega \, \phi} \mathcal{C} \big] = 0
    \>, \qquad \qquad
    \big[ \mathcal{C}^\vee \bigket{\nabla_{-\omega}\, \phi^\vee} = 0
    \>,
\end{align}
where $\phi$ and $ \phi^\vee$ are generic $\brk{n - 1}$-forms, and
$\nabla_{\pm\omega}\>$ is a covariant derivative with connection $\omega$
defined as
\begin{align}
    \nabla_{\pm\omega} \defas \dd \pm \omega
    \>, \quad \text{with} \quad
    \omega \defas \dlog(u) = \sum_{i=1}^n \widehat{\omega}_i \, \dd z_i \>,
    \quad \text{and} \quad
    \widehat{\omega}_i \defas \partial_{z_i} \log(u) \>.
    \label{eq:omega_def}
\end{align}
As a result, $\langle \phiL \vert$ and $\vert \phiR \rangle$  are elements of
the $n$\textsuperscript{th} relative cohomology group $\rmH^{n}$ and its dual
$\rmH^{\vee \, n}$, which we write, following the notation of
\cite{matsumoto2018relative,Caron-Huot:2021iev,Caron-Huot:2021xqj}, as
\begin{eqnarray}
    \bra{\phiL} \in \rmH^n \defas H^n(T,D,\nabla_\omega)
    \>,
    \qquad
    \ket{\phiR} \in \rmH^{\vee \, n} \defas H^n(T^\vee,D^\vee,\nabla_{-\omega})
    \>.
    \label{eq:cohomology_groups}
\end{eqnarray}
The ambient spaces are the Zariski-open subsets of the complex $n$-dimensional space
\begin{eqnarray}
    T = \mathbb{C}^n \setminus\mathcal{P_\omega} \setminus D^\vee
    \>,
    \qquad \qquad
    T^\vee = \mathbb{C}^n \setminus \mathcal{P_\omega} \setminus D
    \>,
    \label{eq:Tdef}
\end{eqnarray}
where the polar set $\mathcal{P_\omega} = V(\prod_i B_i)$, is the vanishing
locus $V$ of the polynomial factors~\eqref{eq:integraldef}, and $D,D^\vee$ are relative boundaries,
which are sets of singularities not regulated by the twist.

Analogously, the integration domains ${\cal C}$ and ${\cal C}^\vee$are elements of
the relative $n$\textsuperscript{th} homology group $H_n$ and its dual
$H_n^\vee$. De Rham's theorem ensures that the four vector spaces are
isomorphic, hence they have the same finite dimension, which can be
evaluated as:
\begin{align}
    \nu
    &= \Dim\left( \rmH^{n} \right)
     = \Dim\left( \rmH^{\vee n} \right)
     = \Dim\left( \rmH_{n} \right)\nonumber
     = \Dim\left( \rmH_{n}^\vee \right)\nonumber \\
    &= \text{number of zeros of $\dlog\brk{u_\rho}$}
    \>.
    \label{eq:leepom}
\end{align}
Assuming $D=
V(D_1\cdot\ldots \cdot D_{n_D})$ and $D^\vee= V(D^\vee_1\cdot\ldots \cdot
D^\vee_{n_D^\vee})$, with $n_D$ and $n_D^\vee$ number of relative surfaces,
$u_\rho$ is the fully regulated twist, defined as:
\begin{eqnarray}
    u_\rho = u(z) \prod_{i=1}^{n_{D}}\bigl(D_i\bigr)^{\rho_i}
    \prod_{j=1}^{n_{D}^\vee}\bigl(D^{\vee }_j\bigr)^{\tau_j}
    \>,
    \label{eq:twist_reg}
\end{eqnarray}
where $\rho_i$ and $\tau_j$ are analytic regulators. The
dimension~\eqref{eq:leepom} is also related to the number of critical points of
the Morse height function $\log (\abs{u_\rho})$, see~\cite{Lee:2013hzt}.

Besides the two pairing that define the integrals $\langle \phiL \vert
\mathcal{C} ]$ and the dual $[ \mathcal{C}^\vee | \phiR \rangle  $, it is
possible to define the other two quantities: \textit{intersection numbers} for
twisted cycles $[\mathcal{C}^\vee \,\vert\,  \mathcal{C}]$ and twisted cocycles
$\vev{\phiL \,\vert\, \phiR}$. The latter is going to be the main subject of this
work.

\subsection{Linear relations}
The finite dimensionality of $\rmH^{n}$ and $\rmH^{\vee n}\>$, ensures the
existence of bases
$\brc{\bra{e_i}}_{i=1}^\nu$ and $\brc{\ket{ h_i}}_{i=1}^{\nu}$
belonging respectively to $\rmH^{n}$ and $\rmH^{\vee n}$, from
which it is possible to construct square matrix of intersection numbers called
the \textbf{C}-matrix or \textit{metric}
\begin{align}
    \Cmat_{ij} \defas \vev{e_i \,|\, {h}_j}
    \>.
    \label{eq:firstCdef}
\end{align}
The fact that the rank of this matrix is $\nu$ means that the intersection numbers
define a non-degenerate scalar product. Moreover, any arbitrary cocycle
$\bra{\phiL}$ can be decomposed in terms of the basis
elements~\cite{Mastrolia:2018uzb, Frellesvig:2019kgj} via the \textit{master
decomposition formula}
\begin{align}
    \langle \phiL |
    =
    \sum_{i=1}^{\nu} c_i \> \bra{{e}_i}
    \>,
    \qquad\quad
    \text{with}
    \qquad\quad
    c_i = \sum_{j=1}^\nu\langle {\phiL} \,|\, {h}_j \rangle (\Cmat^{-1})_{ji}
    \>.
    \label{eq:masterdecomposition}
\end{align}
Consequently, any twisted-period integral $I$ can be decomposed in terms of a
basis of \textit{master integrals} (MIs) $\mathcal{I}_{i}$ as
\begin{align}
    I = \langle \phiL | \mathcal{C} ]
    \>
    = \sum_{i=1}^{\nu} c_i \> \bigbra{e_i} \mathcal{C} \big]
    = \sum_{i=1}^{\nu} c_i \> \mathcal{I}_i
    \>,
    \label{eq:FI_decomposition}
\end{align}
with the coefficients of the decomposition $c_i$ given by
eq.~\eqref{eq:masterdecomposition} and
$\mathcal{I}_i = \bigbra{e_i} \mathcal{C} \big]$.
\subsection{1-form intersection numbers}
In the case of 1-forms, intersection numbers between twisted cocycles of the type:
\begin{eqnarray}
    \phiL \in H^1 (T,D;\nabla_\omega) \quad \text{and} \quad
    \phiR \in H^1 (T^\vee,D^\vee;\nabla_{-\omega})
\end{eqnarray}
are defined as integrals of the product of a representative form and its dual,
after an appropriate regularisation procedure:
\begin{align}
    \langle \phiL \,|\, \phiR \rangle \defas
    \frac{1}{2 \pi i} \int_{\cX} \! \iota(\phiL) \wedge \iota^\vee(\phiR)
    \>.
    \label{eq:interexdefiota}
\end{align}
In this case $\cX = T\cap T^\vee$, and the regulators act on the forms
schematically\footnote{
    Strictly speaking, the local solutions $\psi$ appearing in
    eq.~\eqref{eq:iota_univariate} should bear an additional index $p$
    corresponding to the points they are localized at.
    In the rest of this review we will avoid it to simplify the
    notation, however see~\appref{app:polydiv} for more details.
} as:
\begin{align}
    \iota(\phiL)& \defas \phiL - \sum_{p\in  \Poles_\omega\cup D }\nabla_{\!  \omega}  [(1 - \theta_{z,p}) \psiL] \>,  \nonumber \\
    \iota^\vee(\phiR) &\defas \phiR - \sum_{p\in  \Poles_\omega\cup D^\vee }\nabla_{\! -\omega} [(1 - \theta_{z,p}) \psiR]
    \>,
    \label{eq:iota_univariate}
\end{align}
with the Heaviside functions defined as:
\begin{align}
    \theta_{z,p} \defas \theta(|z {-} p| - \epsilon)\ ,
    \label{eq:hthetadefs}
\end{align}
where we introduced a small real parameter $0 < \epsilon \ll 1$.
Above the $\psiL$ and $\psiR$ are the local solutions of the differential equations:
\begin{eqnarray}
     \nabla_{\!  \omega} \psiL = \phiL \>, \qquad   \ \nabla_{\! - \omega} \psiR = \phiR
      \label{eq:psiunivariate}
\end{eqnarray}
on $T$ and $T^\vee$ respectively.
As originally shown in \cite{matsumoto2018relative},
intersection numbers of twisted $1$-forms with integration variable $z$ can be
evaluated as: 
\begin{align}
    \vev{\phiL \,|\, \phiR} =
    - \sum_{p\in \mathcal{P}_\omega} \Res_{z=p} (\psi^\vee \, \phiL) - \sum_{p\in D^\vee} \Res_{z=p} (\psi^\vee \, \phiL) +  \sum_{p\in D} \Res_{z=p} (\psi \, \phiR)
    \>.
    \label{eq:univariate}
\end{align}

For the computation of the intersection numbers
it is sufficient to find the {\it local} solutions of
eq.~\eqref{eq:psiunivariate} around the poles $\Poles_{\omega}$ \brk{as well as
the corresponding relative set $D$ or $D^\vee$} contributing to the residue
formula~\eqref{eq:univariate}.
As discussed\footnote{
    See also~\cite{Cacciatori:2021nli} for a nice review.
} in \cite{Weinzierl:2020xyy,Fontana:2023amt}, the sum over the poles
$\mathcal{P}_\omega$ can be computed as the {\it global residue}
\begin{eqnarray}
    \sum_{p\in \mathcal{P}_\omega} \Res_{z=p} (\psi^\vee \, \phiL)\  = \  \Res_{\langle \Baikov \rangle} ( \psiR \, \phiL )
    \>,
\end{eqnarray}
where $\Baikov\brk{z} $ is a univariate polynomial
vanishing on the polar set \brk{see~\appref{app:code} for more details on its
choice}
\begin{align}
    \Poles_\omega \subseteq V\brk{\Baikov}
    \>.
\end{align}
The vector of local solutions~\eqref{eq:psiunivariate} can be viewed\footnote{
    See~\appref{app:polydiv} for details of this transformation.
} as an element of the quotient ring~\cite{Brunello:2023rpq}
$\cQ \defas \Field\sbrk{z}/\vev{\Baikov}$, where
$\Field\sbrk{z}$ is the space of polynomials in the variable $z$ with
coefficients in the field $\Field$,
$\vev{\Baikov}$ is the ideal generated by the polynomial $\Baikov\brk{z} -
\beta$ deformed by the symbolic parameter $\beta$.

In our applications, $D = \varnothing$ and only simple poles in $D^\vee$ will
appear, hence the last term on the RHS of eq.~\eqref{eq:univariate} will not
contribute to the intersection number.
{This is always the case for FIs in Baikov representation, since higher
order poles can be mapped into derivatives of the twist via integration by
parts}.

Moreover, we will distinguish the case in which $\varphi^\vee$ is a
boundary-supported form, meaning a form living on $D^\vee$ as defined below, or not.

\paragraph{Regulated case}
If $\varphi^\vee$ is not a boundary-supported form, only the first term on the
RHS of eq.~\eqref{eq:univariate} will contribute to the intersection number,
so it becomes:
\begin{eqnarray}
     \vev{\phiL \,|\, \phiR} & =& -\Res_{\vev{\Baikov}} \Bigbrk{
        \phiL \> \psi
    }
    \>,
\end{eqnarray}
where from now on we omit the $\vee$ sign from the solution.
Therefore, working in $\mathcal{Q}$, the intersection number can be evaluated
by solving the system of equations:
\begin{alignat}{2}
    \vev{\phiL \,|\, \phiR}
    +
    \Res_{\vev{\Baikov}} \Bigbrk{
        \phiL \> \psi
    }  & = &
    \> & 0
    \>,
    \label{eq:global_residue_univariate}
    \\[2mm]
    \Bigfloor{
        \widehat{\nabla}_{-\omega} \, \psi
        - \hatPhiR
    }_{\vev{\Baikov}} & = &
             \> & 0
    \>,
    \label{eq:system_modular_univariate}
\end{alignat}
where the bracket $\bigfloor{\>}_{\vev{\Baikov}}$ stands for the 
remainder of the polynomial division w.r.t. ${\Baikov}$, and
\begin{eqnarray}
    \widehat{\nabla}_{-\omega}
    \equiv
        \bigbrk{\partial_{z} \Baikov} \, \partial_\beta
        - \widehat{\omega} \>
        + \partial_{z}
    \ .
\end{eqnarray}
The solution can be obtained by the following ansatz~\cite{Fontana:2023amt}:
\begin{align}
    \psi\brk{\beta, z} =
    \sum_{a = 0}^{\kappa - 1}
    \>
    \sum_{n \in \Integers}
    \>
    z^a \, \beta^n
    \>
    \psi_{an}
    \>,
    \quad
    \text{with $\kappa \defas \mathrm{deg}\brk{\Baikov}$}
    \>,
    \label{eq:ansatz_univariate}
\end{align}
where the Laurent series expansion coefficients $\psi_{an}$
vanish for small enough $n$.
The global residue can then be evaluated as:
\begin{align}
    \vev{\phiL \,|\, \phiR} = - \frac{1}{\ell_c} \>
    \Bigfloor{{\hatPhiL} \psiL}_{\vev{\Baikov}}
    \Bigg\vert_{z^{\kappa-1}\beta^{-1}}
     \>,
\end{align}
that is the coefficient of the term $z^{\kappa - 1} \> \beta^{-1}$ divided by
the leading coefficient $\ell_c$ of the $\Baikov$ polynomial.

\paragraph{Relative case}
Assuming $z\in D^\vee$ to be a non-regulated boundary, we allow the existence of boundary-supported forms in $H^{\vee \, 1}$ defined via the Leray coboundary
operation \cite{matsumoto2018relative,Caron-Huot:2021xqj,Caron-Huot:2021iev,Brunello:2023rpq}
\begin{align}
    \delta_{z}(\phi^\vee (z)) := \frac{u(z)}{u(0)} \, \dd \theta_{z,0} \, \phi^\vee (z)
    \>,
    \label{eq:deltadef}
\end{align}
where  $\phi^\vee (z)$ is a generic univariate function regular at $z=0$ and the Heaviside function is defined as in eq.~\eqref{eq:hthetadefs}. In this case, only the second term in eq.~\eqref{eq:univariate} will contribute, and
intersection numbers localise on the non-regulated pole as:
\begin{align}
    \braket{ \phiL }{  \delta_z(\phi^\vee (z)) }
    := \Res_{z = 0} \left( \frac{u(z)}{u(0)} \phiL \ \phi^\vee (z)\right)
    \>.
    \label{eq:relativeunivariate}
\end{align}
Let us remark, that an analogous left form can be introduced for $z \in D$,
however it will not be used in the applications considered in this work.

\subsection{$n$-form intersection numbers}
\label{sec:fibration}
Intersection numbers for $n$-forms can be evaluated using the
\textit{fibration}-based approach introduced in
\cite{Ohara98intersectionnumbers, Mizera:2019gea, Frellesvig:2019uqt,Frellesvig:2020qot}
and discussed more recently in~\cite{Caron-Huot:2021iev,Caron-Huot:2021xqj,Fontana:2023amt,Brunello:2023rpq},
in which the integration variables are considered one at a time. Once an order
of variables has been fixed\footnote{
    In principle, any order of variables is valid here. In practice, it is chosen
    to minimize the number of independent basis elements in each layer of the
    fibration.
}, for example  $\{z_n,\ldots, z_1\}$, where on the
right appears the innermost variable, multivariate intersection numbers can be
evaluated recursively.

Let us focus on the $m^{\text{th}}$ fiber, assuming that all the $\brk{m -
1}$-variate building blocks are known. Specifically, given the bases
$\{e\supbbf{m-1}_i\}_{i=1}^{\nu_{m-1}}$ and
$\{h\supbbf{m-1}_i\}_{i=1}^{\nu_{m-1}}$ of $H^{m-1}$ and its dual $H^{\vee\
m-1}$, respectively, we define the $\Cmat$-matrix at the $(m-1)^\text{th}$
layer as:
\begin{align}
    \Cmat^{(m-1)}_{ij}
    &\defas
    \vev{
        e\supbbf{m-1}_i \,|\, h\supbbf{m-1}_j
    }
    \>.
\end{align}
We introduce the covariant derivatives on the $m^\text{th}$ fibration layer
\begin{eqnarray}
    \bigbrk{\nabla_{\Omega^{(m)}}}_{ij} \defas \delta_{ij} \dd_{z_m} + \Omega^{(m)}_{ji}
    \>, \quad
    \bigbrk{\nabla_{\Omega^{\vee  (m)}}}_{ij} \defas \delta_{ij} \dd_{z_m} + \Omega^{\vee (m)}_{ij}
    \>,
\end{eqnarray}
where $\delta_{ij}$ denotes the Kronecker delta function.
The matrix-valued
connection operators
$\Omega^{(m)}$ and
$\Omega^{\vee (m)}$
are defined by the following systems of differential equations,
fulfilled by the bases and dual bases elements
\begin{eqnarray}
    \dd_{z_m} \> \Bigbra{e_i^{(\mathbf{m-1})}} \>
    = \sum_{j=1}^{\nu_{m-1}}
    \Omega^{(m)}_{ij} \>
    \Bigbra{e_j^{(\mathbf{m-1})}}
    \>, \quad
    \dd_{z_m} \Bigket{h_i^{(\mathbf{m-1})}} \>
    = -\sum_{j=1}^{\nu_{m-1}}
    \Omega^{\vee(m)}_{ji} \>
    \Bigket{h_j^{(\mathbf{m-1})}}
    \>,
\end{eqnarray}
and computed in terms of intersection numbers, from the master decomposition formulae
\begin{align}
    \widehat{\Omega}^{(m)}_{ij}
    &=\sum_{k=1}^{\nu_{m-1}}
    \Bigvev{
        \brk{
            \partial_{z_m} + \widehat{\omega}_m
        }e\supbbf{m - 1}_i
        \,|\,
         h\supbbf{m - 1}_k
    }\>
    \bigbrk{\Cmat_{(m-1)}^{-1}}_{kj}
    \> \>,
    \nonumber
    \\
    \widehat{\Omega}^{\vee (m)\, }_{ji}
    & = -\sum_{k=1}^{\nu_{m-1}}
    \bigbrk{\Cmat_{(m-1)}^{-1}}_{ik}
    \>
    \Bigvev{
        e\supbbf{m - 1}_k
        \,|\,
        \brk{
            \partial_{z_m} - \widehat{\omega}_m
        } h\supbbf{m - 1}_j
    }
    \>.
    \label{eq:psimultivariatedual}
\end{align}
Differential $m$-forms can be rewritten as:
\begin{align}
    \phiL^{\brkbf{m}} = \sum_{i = 1}^{\nu_{m - 1}} \phiL^{\brk{m}}_i
    \wedge
    e\supbbf{m - 1}_i
    \>, \qquad \qquad
    \phiR^{\brkbf{m}} = \sum_{i = 1}^{\nu_{m - 1}} \phiR^{\brk{m}}_i
    \wedge
    h\supbbf{m - 1}_i
    \>,
    \label{eq:projection}
\end{align}
where the projection coefficients $\phiL^{\brk{m}}_i$ and $\phiR^{\brk{m}}_i$
are indeed 1-forms and are computed through the master decomposition formulae
\begin{align}
    \projPhi^{(m)}_{i}
    &=
    \sum_{j=1}^{\nu_{m-1}} \vev{
        \phiL^{\brkbf{m}} \,|\, h\supbbf{m - 1}_j
    }
    \,
    ( \Cmat_{(m-1)}^{-1} )_{ji}
    \> \>,
    \nonumber
    \\
    \projPhi^{\vee (m)}_{i}
    &=
    \sum_{j=1}^{\nu_{m-1}}
    ( \Cmat_{(m-1)}^{-1} )_{ij}
    \,
    \vev{
        e^{(\mathbf{m-1})}_j \,|\, \phiR^{\brkbf{m}}
    }
    \> .
    \label{eq:projections}
\end{align}

As a starting point for the discussion of the novel ideas presented in this work,
we conveniently cast them~\cite{Ohara98intersectionnumbers, Mizera:2019gea} in vector-valued $1$-forms of length $\nu_{m-1}$
\begin{align}
    \phiL^{\brk{m}} =
    \begin{bNiceMatrix}[margin]
        \phiL^{\brk{m}}_1 & \Ldots[color = gr] & \phiL^{\brk{m}}_{\nu_{m - 1}}
    \end{bNiceMatrix}\tr
    &\in H^1\brk{T^{(m)}, D^{(m)}; \nabla_{\Omega^{(m)}}}
    \>,
    \nonumber
    \\
    \phiR^{\brk{m}} =
    \begin{bNiceMatrix}[margin]
        \phiR^{\brk{m}}_1 & \Ldots[color = gr] & \phiR^{\brk{m}}_{\nu_{m - 1}}
    \end{bNiceMatrix}\tr
    &\in H^1 (T^{\vee (m)},D^{\vee (m)};\nabla_{\Omega^{\vee  (m)}})
    \>,
    \label{eq:vector_valued_forms}
\end{align}
where:
\begin{eqnarray}
    T^{(m)} = \mathbb{C} \setminus\mathcal{P}_{\Omega^{(m)}} \setminus D^{\vee(m)}
    \>,
    \qquad \qquad
    T^\vee = \mathbb{C} \setminus \mathcal{P}_{\Omega^{(m)}} \setminus D^{(m)}
    \>,
\end{eqnarray}
and the polar set $\Poles_{\Omega^{(m)}}$ is the set of singularities of
$\Omega^{(m)}$, while $D^{(m)}$ and $D^{\vee(m)}$ are the relative boundaries at layer $m$.
Hence, intersection numbers for differential $m$-forms can be evaluated as:
\begin{align}
     \vev{\phiL^{\brkbf{m}} \,|\, \phiR^{\brkbf{m}}}  =
     & -\sum_{p\in \mathcal{P}_{\Omega^{(m)}}} \Res_{z_m=p} \Bigbrk{
        \sum_{i=1}^{\nu_{m-1}}\vev{
            \phiL^{\brkbf{m}}\vert h\supbbf{m - 1}_i
        }
        \,
        \psiR_{i}^{(m)}
        } \nonumber\\
    &
    - \sum_{p\in D^{\vee(m)}}\Res_{z_m=p}\Bigbrk{
        \sum_{i=1}^{\nu_{m-1}}\vev{
            \phiL^{\brkbf{m}}\vert h\supbbf{m - 1}_i
        }
        \,
        \psiR_{i}^{(m)}
        }\nonumber \\
    &
    + \sum_{p\in D^{(m)}}\Res_{z_m=p}\Bigbrk{
    \sum_{i=1}^{\nu_{m-1}}\psiL_{i}^{(m)}
        \vev{
        e\supbbf{m - 1}_i
            \vert \phiR^{\brkbf{m}}
        }
        }
    \>,
    \label{eq:multivariate}
\end{align}
where $\psiL$ and $\psiR$ are solutions to the systems of differential equations:
\begin{align}
    \nabla_{\Omega^{(m)}}\,
    \psiL^{(m)} = \phiL^{(m)}
    \>, \qquad
    \nabla_{\Omega^{\vee(m)}}\,
    \psiR^{(m)} = \phiR^{(m)}
    \>.
    \quad
    \>
    \label{eq:psimultivariate}
\end{align}
As in the case of 1-forms, it is sufficient to solve the
system~\eqref{eq:psimultivariate} around the set of singularities where the
residues are computed. Being $\Baikov(z_m)$ an univariate polynomial vanishing
on the set of points $\mathcal{P}_{\Omega^{(m)}}\>$, the sum over the poles
$\mathcal{P}_{\Omega^{(m)}}$ can be computed as the global residue:
\begin{eqnarray}
    \sum_{p\in \mathcal{P}_{\Omega^{(m)}}} \Res_{z_m=p} \Bigbrk{\vev{
            \phiL^{\brkbf{m}}\vert h\supbbf{m - 1}
        }
        \cdot
        \psiR^{(m)}}\  = \  \Res_{\langle \Baikov^{(m)} \rangle} \Bigbrk{ \vev{
            \phiL^{\brkbf{m}}\vert h\supbbf{m - 1}
        }
        \cdot
        \psiR^{(m)}
        }
        \>.
\end{eqnarray}

In our applications, $D = \varnothing$ and only simple poles in $D^\vee$ will
appear, hence the last term in the sum in RHS of eq.~\eqref{eq:multivariate}
will not contribute to the intersection number.

\paragraph{Regulated case}
Assuming that $\varphi^{\vee(\mathbf{m})}$ is not a boundary-supported form on
$D^\vee$, only the first term on the RHS of eq.~\eqref{eq:multivariate} will
contribute to the intersection number, and eq.~\eqref{eq:multivariate}
becomes:
\begin{align}
     \vev{\phiL^{\brkbf{m}} \,|\, \phiR^{\brkbf{m}}}  =
     -
     \Res_{\langle \Baikov^{(m)} \rangle} \Bigbrk{ \vev{
            \phiL^{\brkbf{m}}\vert h\supbbf{m - 1}
        }
        \cdot
        \psiL^{(m)}
        }\ .
\end{align}
Working in the quotient ring
$\cQ \defas \Field\sbrk{z_m}/\vev{\Baikov^{(m)}}$, where $\vev{\Baikov^{(m)}}$
denotes the ideal generated by the $\Baikov\brk{z_m} - \beta$ polynomial in the
$z_m$ variable, the system of equations can be
rewritten as:
\begin{alignat}{2}
    \vev{\phiL^{\brkbf{m}} \,|\, \phiR^{\brkbf{m}}}
    +
    \Res_{\langle \Baikov^{(m)}\rangle} \Bigbrk{
        \vev{
            \phiL^{\brkbf{m}}\vert h\supbbf{m - 1}
        }
        \cdot
        \psi\supbrk{m}
    }  & = &
    \> & 0
    \>,
    \label{eq:global_residue}
    \\[2mm]
    \Bigfloor{
        \widehat{\nabla}_{\Omega^{\vee (m)}}
        \psi\supbrk{m}
        - \hatPhiR^{\brk{m}}
    }_{\vev{\Baikov\supbrk{m}}} & = &
                       \> & 0
    \>,
    \label{eq:system_modular}
\end{alignat}
with
\begin{eqnarray}
    \widehat{\nabla}_{\Omega^{\vee (m)}}
    \equiv
\bigbrk{\partial_{z_m} \Baikov\supbrk{m}} \, \partial_\beta
            + \widehat{\Omega}^{\vee \brk{m}}
        + \partial_{z_m} \>,
\end{eqnarray}
where we omitted the cohomology group indices for brevity, and it can be solved
using the ansatz
\begin{align}
    \psi\supbrk{m}_i\brk{\beta, z_m} =
    \sum_{a = 0}^{\kappa - 1}
    \>
    \sum_{n \in \Integers}
    \>
    z_m^a \, \beta^n
    \>
    \psi\supbrk{m}_{ian}
    \quad
    \text{for $i = 1, \ldots, \nu_{m - 1}$}
    \>,
    \label{eq:ansatz_multivariate}
\end{align}
similar to eq.~\eqref{eq:ansatz_univariate}.

\paragraph{Relative case}
Assuming $D^\vee = V\brk{z_1 \cdot \ldots \cdot z_m}$ to be non-regulated boundaries, we allow the existence of multivariate boundary-supported forms
defined via the Leray coboundary operation:
\begin{align}
    \delta_{z_1,\ldots,z_m}(\phi^{\vee(\mathbf{n-m})}) := \frac{u}{u(0)} \bigwedge_{i=1}^m \dd \theta_{z_i,0}\wedge \phi^{\vee(\mathbf{n-m})}
    \>,
    \label{eq:deltamultivariate}
\end{align}
where $\phi^{\vee(\mathbf{n-m})}$ is a generic $\brk{n - m}$-form, the
Heaviside functions are defined in eq.~\eqref{eq:hthetadefs}, and, in
shorthand notation, $u(0) \defas u|_{z_1,\ldots,z_m = 0}\>$.
In this case, intersection numbers of $n$-forms can be evaluated as:
\begin{eqnarray}
    \vev{\phiL^{(\mathbf{n})} | \delta_{z_1,\ldots,z_m} (\phi^{\vee(\mathbf{n-m})}})
    &=&
    \bigvev{\
        \Res_{z_1=0,\ldots,z_m=0} \Bigbrk{
            \tfrac{u}{u(0)} \phiL^{(\mathbf{n})}
        }
        \> \big| \>
        \phi^{\vee(\mathbf{n-m}))
    }}
    \label{eq:relativemultivariate}
    \>,
\end{eqnarray}
where the $\brk{n-m}$-form intersection numbers on the RHS is evaluated using
eq.~\eqref{eq:system_modular}.

\subsection{Feynman integrals as twisted period integrals}
\label{sec:FI_twist}
Feynman Integrals (FIs) can be considered as twisted period integrals. In the
\textit{Baikov representation}~\cite{Baikov:1996iu,Frellesvig:2017aai}, a FI
with $E$ independent external momenta, and $L$ loops can be written in terms of
$N = L\cdot E + L(L+1)/2$ integration variables $z_i\>$, corresponding to the
generalised set of denominators
\begin{align}
    I &= \kappa \int_{\mathcal{C}} \! u \> \phiL= \langle \phiL \vert \mathcal{C}] ,
    \qquad
    u = \Baikov^\gamma,
    \qquad
    \phiL = \frac{\dd^N z}{\prod_{i=1}^N z_i^{a_i}}
    \>,
    \label{eq:baikovdef}
\end{align}
where $\Baikov$ is the Baikov polynomial, which vanishes on the boundary of the
integration domain $\Baikov(\partial \mathcal{C})=0$ guaranteeing that the
integrals~\eqref{eq:baikovdef} obey IBPs.

Let us assume that the first $n$ of the generalised denominators appear as
propagators, $\brc{z_1,\ldots, z_n}$, whereas the remaining integration
variables $\brc{z_{n+1},\ldots, z_N}$ can enter only in the numerator as
irreducible scalar products. The differential $n$-forms we will be dealing with
are of the type:
\begin{eqnarray}
    \phiL = {\hatPhiL}(z) \> \dd^N z
    \>, \qquad
    {\hatPhiL}(z) = \frac{\prod_{j=n+1}^N z_j^{-a_j}}{\prod_{i=1}^n z_i^{a_i}} \ .
\end{eqnarray}
In this case, the definitions for the cohomology group and its dual, given in
eq.~\eqref{eq:cohomology_groups}, specify as:
\begin{align}
    \mathcal{P}_\omega= V(\Baikov)\>, \qquad D= \varnothing \>, \qquad D^\vee = V(z_1\cdot z_2 \cdot\ldots \cdot z_n)\>,
    \label{eq:def_spaces}
\end{align}
and the dimension of the cohomology group $\nu$, hence the number of master
integrals, are given by eq.~\eqref{eq:leepom}, with the regulated
twist~\eqref{eq:twist_reg}. Any FI $I=\langle \phiL\vert \mathcal{C}]$ can be
decomposed in terms of a basis of master integrals (MIs)
$\{\mathcal{I}_{i}\}_{i=1}^\nu\>$, where $\mathcal{I}_i = \langle e_i |
\mathcal{C} \big]$, as in eq.~\eqref{eq:FI_decomposition},
and the coefficients of the decomposition $c_i$ are given by
eq.~\eqref{eq:masterdecomposition}, and the $N$-form intersection numbers are
computed using the fibration approach described in~\secref{sec:fibration}.

We define an integral sector $\mathcal{S}=(\sigma_1,\ldots,\sigma_N)$ as the
set of points $(a_1,\ldots,a_N)\in \mathbb{Z}^N\>$, such that $\sigma_i =
\theta(a_i-1/2)$. We notice that for each sector a subset of relative surfaces
$D^\vee_\mathcal{S}= D^\vee \cap \prod_{i=1}^N z_i^{\sigma_i}$ appears. The
dimension of the cohomology group $\nu$ can be evaluated as sum of the
dimensions of each sector $\nu_\mathcal{S}$
\begin{equation}
    \nu = \sum_\mathcal{S} \nu_\mathcal{S} ,
\end{equation}
where the dimension $\nu_\mathcal{S}$ can be evaluated as:
\begin{eqnarray}
\nu_\mathcal{S} =  \text{number of solutions of } \omega_\mathcal{S}=0\>,
    \qquad  \omega_\mathcal{S} = \dlog(u_\mathcal{S})
    \>,
\end{eqnarray}
with $u_\mathcal{S}$ being the twist of the sector $\mathcal{S}$, obtained from
$u$ by cutting the relative surfaces $D^\vee_{\mathcal{S}}\>$.
In the application discussed in~\secref{sec:pentabox},
we successfully used this algorithm to generate bases in each layer of the fibration
by choosing $\nu_\mathcal{S}$ independent elements in each sector\footnote{
    A proof-of-concept implementation of this algorithm, together with examples
    of its usage, can be found in the following GitHub repository
    \href{https://github.com/GiacomoBrunello/pentabox_decomposition}{\faGithub}.
}.

With this we conclude our overview of the relative twisted cohomology framework
and advance to the tensorial reformulation of the intersection number
computation process.

\section{Tensor structure of intersection numbers}
\label{sec:ctensor}
The fibration method for evaluation of intersection numbers involves solving system of equations,
such as those found in eqs.~\eqref{eq:global_residue_univariate,eq:system_modular_univariate},
for univariate forms, and in eqs.~\eqref{eq:global_residue,eq:system_modular}, for the
multivariate cases.
Matrix calculus provides a powerful approach to reformulating these
computations, offering deeper insight into the mathematical properties and
patterns of intersection numbers while significantly boosting computational
efficiency.

In computational algebraic geometry, \textit{companion
matrices}~\cite{Sturmfels:2002, Cox2015}~\footnote{
    See also~\cite{TelenThesis} for a nice review.
}
allow for the reinterpretation of polynomial division as matrix multiplication.
In this section, we investigate how companion matrices can be used
to develop a novel scheme for computing intersection numbers, ultimately
determining them through matrix operations.

\subsection{The three vector spaces}
Let us fix the fibration layer $m$ and denote by $z$ the corresponding Baikov
variable. In the following we will omit the layer index whenever it does not
cause confusion.
Recall, that the ansatz~\eqref{eq:ansatz_multivariate} is parameterized by 3
indices
\begin{align}
    \psiL^{\brk{m}}_i =
    \>
    \sum_{a \, n}
    \>
    z^a \, \beta^n
    \>
    \psiL_{ian}
    \>,
    \label{eq:ansatz_multivariate_cmats}
\end{align}
which motivates us to distinguish the following 3 vector spaces:
\begin{enumerate}
    \item
        Vector space of $\nu$-dimensional vectors labeled by the first index
        $i = 1, \ldots, \nu$ in eq.~\eqref{eq:ansatz_multivariate_cmats}, namely
        \begin{align}
            \Field^\nu
            \>,
        \end{align}
        closely related to the linear space of vector-valued 1-forms introduced in
        eqs.~\eqref{eq:vector_valued_forms}.
    \item
        The second index $a = 0, \ldots, \kappa - 1$ parameterizes the set of irreducible monomials of
        the ideal $\vev{\Baikov\brk{z} - \beta}$, that is a basis of the quotient ring
        $\cQ$ viewed as a vector space
        \begin{align}
            \cQ
            =
            \Span_{\Field}\bigbrk{{1, \ldots, z^{\kappa - 1}}}
            \>,
            \quad
            \kappa \defas \Deg\bigbrk{\Baikov\brk{z}}
            \>.
            \label{eq:space_quotient}
        \end{align}
        It is then natural to interpret the $\Baikov\brk{z} - \beta = 0$
        equation as a many-to-one coordinate change $z \mapsto \beta$,
        and think of $\cQ$ as a vector bundle of rank $\kappa$ with $\beta$
        parameterizing the base and irreducible monomials forming the basis in
        fibres \brk{see~\appref{app:polydiv} for more details}.
    \item
        Finally, the last index $n \in \Integers$ runs over the powers of the
        variable $\beta$ that appear in the Laurent series expansion
        \begin{align}
            \cL = \Span_{\Field}\bigbrk{
                \ldots, \, \beta^{-1}, \, \beta^0, \, \beta^1, \, \ldots
            }
            \>.
            \label{eq:space_laurent}
        \end{align}
        The $z \mapsto \beta$ coordinate change introduced above effectively
        glues all the roots of the $\Baikov\brk{z}$ polynomial in the $z$-plane
        to the origin of the $\beta$-plane.
        The space~\eqref{eq:space_laurent} captures the local behavior of
        functions at this point.
\end{enumerate}
The solution $\psiL$ to the differential equation
system~\eqref{eq:ansatz_multivariate_cmats} belongs to the tensor product
of these three spaces:
\begin{align}
    \psiL^{\brk{m}} \in \Field^{\nu} \otimes \cQ \otimes \cL
    \>.
    \label{eq:tensor_space}
\end{align}
In the following we will represent the differential equation
system~\eqref{eq:system_modular_univariate, eq:system_modular} as a linear
operator $\ctensor{\nabla}$ acting on the space~\eqref{eq:tensor_space}
\begin{align}
    \ctensor{\nabla_{\Omega^\vee}} \cdot \psiL^{\brk{m}} - \hatPhiR^{\brk{m}} = 0
    \>.
    \label{eq:system_modular_tensor}
\end{align}
This formulation allows the solution to be obtained with linear algebra methods.

\subsubsection{Polynomial companion matrices}
\label{sssec:cmat}
For the ideal generated by a monic\footnote{
    Here we normalized the $B_\kappa$ coefficient to 1 to avoid clutter
    in the subsequent formulae.
} degree $\kappa$ polynomial
\begin{align}
    \vev{\Baikov}
    \equiv
    \vev{\Baikov\brk{z} - \beta}
    =
    \vev{
        b_0 - \beta + z \, b_1 + \ldots
        + z^{\kappa - 1} \, b_{\kappa - 1}
        + z^{\kappa}
    }
    \>,
    \label{eq:poly_def}
\end{align}
we choose as a basis of the quotient ring the list of monomials shown in
eq.~\eqref{eq:space_quotient}.
In the monomial basis of eq.~\eqref{eq:space_quotient}, the companion matrix
representation for the basic multiplication and differentiation
operators read
\begin{align}
    Q_{z}
    \defas
    \begin{NiceArray}{ccc ccl}[right-margin=8mm, left-margin=6mm]
        \gr{0} & & & \hspace*{5mm} & & - {b_{0}} + \beta
        \\
        1 & \gr{0} & & & & -{b_{1}}
        \\
        & 1 & \gr{0} & & & -{b_{2}}
        \\[6mm]
        & & \Ddots[color=black] & \Ddots[color=gr] & & \Vdots[color = gr]
        \\
        & & & 1 & \gr{0} & -{b_{\kappa - 2}}
        \\
        & & & & 1 & -{b_{\kappa - 1}}
        \\
        \CodeAfter
            \SubMatrix[{1-1}{6-6}][xshift=2mm]
            \tikz \draw[<->, gr, shorten <> = .5em]
                (1-|7.5) ++(6mm, 0)
                -- ($(7-|7.5) + (6mm, 0)$)
                node [midway, fill=white] {\scriptsize$\kappa$}
            ;
            \tikz \draw[<->, gr, shorten <> = .5em]
                (6.5-|1) ++(0, -.5)
                -- ($(6.5-|7.5) + (0, -.5)$)
                node [midway, fill=white] {\scriptsize$\kappa$}
            ;
    \end{NiceArray}
    \hspace{1.5cm}
    Q_{\partial_z}
    \defas
    \begin{NiceArray}{ccc ccl}[right-margin=8mm, left-margin=4mm]
        \CodeBefore
        \Body
        \gr{0} & 1 & & & \hspace*{5mm} &
        \\
        & \gr{0} & 2 & & &
        \\
        & & \gr{0} & 3 & &
        \\[6mm]
        & & & \Ddots[color=gr] & \Ddots[color=black, shorten-end=-2mm] &
        \\
        & & & & & {\kappa - 1}
        \\
        & & & & & \gr{0}
        \CodeAfter
            \SubMatrix[{1-1}{6-6}][xshift=2mm]
            \tikz \draw[<->, gr, shorten <> = .5em]
                (1-|7.5) ++(6mm, 0)
                -- ($(7-|7.5) + (6mm, 0)$)
                node [midway, fill=white] {\scriptsize$\kappa$}
            ;
            \tikz \draw[<->, gr, shorten <> = .5em]
                (6.5-|1) ++(0, -.5)
                -- ($(6.5-|7.5) + (0, -.5)$)
                node [midway, fill=white] {\scriptsize$\kappa$}
            ;
    \end{NiceArray}
    \label{eq:cmats_fin}
\end{align}
\vspace{2mm}

\noindent
where we explicitly showed the $0$ entries on the main diagonal for the
reader's convenience.

\subsubsection{Series companion matrices}
To operate on Laurent series expansions in the $\beta$ variable, we employ the
infinite matrix representation of the Weyl algebra composed of the two
operators
\begin{align}
    L_{\beta}
    \defas
        \NiceMatrixOptions{code-for-first-col = \scriptstyle}
        \begin{bNiceArray}{c ccc ccc ccc}[margin, nullify-dots, first-col]
            \CodeBefore
            \Body
            & \phantom{\gr{0}} & & & & & & & &
        \\
            & \phantom{1} & & & & & & & &
        \\
            & & \Ddots[shorten-end=-1mm] & \Ddots[color = gr, draw-first] & & & & & &
        \\
            \gr{-1 \rightarrow}
            & & & 1 & \gr{0} & & & & & &
        \\
            \gr{0 \rightarrow}
            & & & & 1 & \gr{0} & & & &
        \\
            \gr{1 \rightarrow}
            & & & & & 1 & \gr{0} & & &
        \\
            \gr{2 \rightarrow}
            & & & & & & 1 & \gr{0} & &
        \\
            & & & & & & & \Ddots & \Ddots[color = gr, shorten-end=0mm] &
        \\
            & & & & & & & & \phantom{1} & \phantom{\gr{0}}
        \CodeAfter
            \tikz \draw[
                gr,
                line cap = round,
                line width = 1.06pt,
                dash pattern= on 0pt off .18cm
            ]
                (1|-1.5) ++(-.5, 0)
                -- ++($(1|-4.5) - (1|-1.5) + (0, .2)$)
                ++($(1|-8) - (1|-4.5) - (0, .2)$)
                -- ++($(1|-8.5) - (1|-7) - (0, .2)$)
            ;
            \tikz \fill[
                fill = gr, fill opacity = .1,
                rounded corners
            ]
                (1|-4) ++(0, -.07) -- ++($(11|-4) - (1|-4)$) |- ($(5-|1) + (0, .07)$) -- cycle
            ;
    \end{bNiceArray}
    \hspace{1.5cm}
    L_{\partial_\beta}
    \defas
        \begin{bNiceArray}{c ccc ccc ccc}[margin, nullify-dots, first-col]
            & \phantom{\gr{0}} & \phantom{0} & & & & & & &
        \\
            & & & & & & & & &
        \\
            & & & \Ddots[color = gr] & \Ddots & & & & &
        \\
            \gr{-1 \rightarrow}
            & & & & \gr{0} & 0 & & & &
        \\
            \gr{0 \rightarrow}
            & & & & & \gr{0} & 1 & & &
        \\
            \gr{1 \rightarrow}
            & & & & & & \gr{0} & 2 & &
        \\
            \gr{2 \rightarrow}
            & & & & & & & \gr{0} & 3 &
        \\
            & & & & & & & & \Ddots[color = gr] & \Ddots
        \\
            & & & & & & & & & \phantom{\gr{0}}
        \CodeAfter
            \tikz \draw[
                gr,
                line cap = round,
                line width = 1.06pt,
                dash pattern= on 0pt off .18cm
            ]
                (1|-1.5) ++(-.5, 0)
                -- ++($(1|-4.5) - (1|-1.5) + (0, .2)$)
                ++($(1|-8) - (1|-4.5) - (0, .2)$)
                -- ++($(1|-9.5) - (1|-8) - (0, .2)$)
            ;
            \tikz \fill[
                fill = gr, fill opacity = .1,
                rounded corners
            ]
                (1|-4) ++(0, -.07) -- ++($(10|-4) - (1|-4)$) |- ($(5-|1) + (0, .07)$) -- cycle
            ;
    \end{bNiceArray}
    \label{eq:cmats_weyl_infinite}
\end{align}
\vspace{2mm}

In practice, however, only a finite number of the ans\"atze
coefficients~\eqref{eq:ansatz_univariate, eq:ansatz_multivariate}
contribute to a given intersection number, suggesting that the infinite
matrix representation~\eqref{eq:cmats_weyl_infinite} should be somehow
restricted so that the matrices become finite.

Let us denote the leading exponents of the Laurent series expansions in $\beta$ of the
building blocks of the main linear system~\eqref{eq:system_modular_univariate, eq:system_modular} as
\begin{align}
    \mathrm{min}_\beta\bigbrk{\widehat{\Omega}^{\vee \brk{m}}} = -1
    \>, \quad
    \mathrm{min}_\beta\bigbrk{
        \vev{
            \phiL^{\brkbf{m}}\vert h\supbbf{m - 1}
        }
    }
    = \muL
    \>, \quad
    \mathrm{min}_\beta\bigbrk{\hatPhiR^{\brk{m}}} = \muR \>,
    \label{eq:omega_expansion}
\end{align}
where we would like to emphasize the simple pole condition for the
connection matrix, which we always can satisfy with an appropriate choice of
the $\Baikov\brk{z}$ polynomial~\eqref{eq:poly_def} \brk{see~\appref{app:code}
for more details}.
Now we may introduce the restricted analog%
\footnote{
    Strictly speaking, finite matrices do not form a representation of a Weyl
    algebra. For example, the finite matrix $L_{\beta}$ shown in
    eq.~\eqref{eq:cmats_weyl} does not have a well-defined inverse: the
    possible candidate $L_{1 / \beta}$ does not give a pure identity matrix in
    the products $L_{\beta} \cdot L_{1 / \beta} = \mathrm{diag}\brk{0, 1,
    \ldots, 1}$ or $L_{1 / \beta} \cdot L_{\beta} = \mathrm{diag}\brk{1,
    \ldots, 1, 0}$.
    Nevertheless, in practice we may use the rules~\eqref{eq:cmats_weyl} after a
    \brk{symbolic} series expansion in the $\beta \to 0$ limit.

} of the infinite representation~\eqref{eq:cmats_weyl_infinite} reading
\begin{align}
    L_{\beta}
    \defas
        \NiceMatrixOptions{code-for-first-col = \scriptstyle}
        \begin{bNiceArray}{c ccc ccc ccc c}[margin, nullify-dots, first-col]
            \CodeBefore
            \Body
            \gr{\muL + 1 \rightarrow}
            & \gr{0} & & & & & & & & &
        \\
            & 1 & \gr{0} & & & & & & & &
        \\
            & & \Ddots & \Ddots[color = gr] & & & & & & &
        \\
            \gr{-1 \rightarrow}
            & & & 1 & \gr{0} & & & & & & &
        \\
            \gr{0 \rightarrow}
            & & & & 1 & \gr{0} & & & & &
        \\
            \gr{1 \rightarrow}
            & & & & & 1 & \gr{0} & & & &
        \\
            & & & & & & \Ddots & \Ddots[color = gr, shorten-end=0mm] & & &
        \\
            & & & & & & & & & \gr{0} &
        \\
            \gr{-\muR \rightarrow}
            & & & & & & & & & 1 & \gr{0}
        \CodeAfter
            \tikz \draw[
                gr,
                line cap = round,
                line width = 1.06pt,
                dash pattern= on 0pt off .18cm
            ]
                (1|-2) ++(-.5, 0)
                -- ++($(1|-4.5) - (1|-2) + (0, .2)$)
                ++($(1|-7) - (1|-4.5) - (0, .2)$)
                -- ++($(1|-9.5) - (1|-7) + (0, .2)$)
            ;
            \tikz \draw[<->, gr, shorten <> = .5em]
                (1-|11.5) ++(6mm, 0)
                -- ($(10-|11.5) + (6mm, 0)$)
                node [midway, fill=white, rotate=90] {\scriptsize$-\muR - \muL + 1$}
            ;
            \tikz \draw[<->, gr, shorten <> = .2em]
                (9.5-|1) ++(0, -.5)
                -- ($(9.5-|11) + (0, -.5)$)
                node [midway, fill=white] {\scriptsize$-\muR - \muL + 1$}
            ;
            \tikz \fill[
                fill = gr, fill opacity = .1,
                rounded corners
            ]
                (1|-4) ++(0, -.07) -- ++($(11|-4) - (1|-4)$) |- ($(5-|1) + (0, .07)$) -- cycle
            ;
    \end{bNiceArray}
    \hspace{1.5cm}
    L_{\partial_\beta}
    \defas
        \begin{bNiceArray}{c ccc ccc clc}[margin, nullify-dots, first-col]
            \gr{\muL + 1 \rightarrow}
            & \gr{0} & \muL & & & & & & &
        \\
            & & \gr{0} & & & & & & &
        \\
            & & & \Ddots[color = gr, shorten-end=-1mm] & \Ddots[draw-first] & & & & &
        \\
            \gr{-1 \rightarrow}
            & & & & \gr{0} & 0 & & & &
        \\
            \gr{0 \rightarrow}
            & & & & & \gr{0} & 1 & & &
        \\
            \gr{1 \rightarrow}
            & & & & & & \gr{0} & 2 & &
        \\
            & & & & & & & \Ddots[color = gr] & \Ddots &
        \\
            & & & & & & & & \gr{0} & -\muR
        \\
            \gr{-\muR \rightarrow}
            & & & & & & & & & \gr{0}
        \CodeAfter
            \tikz \draw[
                gr,
                line cap = round,
                line width = 1.06pt,
                dash pattern= on 0pt off .18cm
            ]
                (1|-2) ++(-.5, 0)
                -- ++($(1|-4.5) - (1|-2) + (0, .2)$)
                ++($(1|-7) - (1|-4.5) - (0, .2)$)
                -- ++($(1|-9.5) - (1|-7) + (0, .2)$)
            ;
            \tikz \draw[<->, gr, shorten <> = .5em]
                (1-|10.5) ++(6mm, 0)
                -- ($(10-|10.5) + (6mm, 0)$)
                node [midway, fill=white, rotate=90] {\scriptsize$-\muR - \muL + 1$}
            ;
            \tikz \draw[<->, gr, shorten <> = .2em]
                (9.5-|1) ++(0, -.5)
                -- ($(9.5-|10) + (0, -.5)$)
                node [midway, fill=white] {\scriptsize$-\muR - \muL + 1$}
            ;
            \tikz \fill[
                fill = gr, fill opacity = .1,
                rounded corners
            ]
                (1|-4) ++(0, -.07) -- ++($(10|-4) - (1|-4)$) |- ($(5-|1) + (0, .07)$) -- cycle
            ;
    \end{bNiceArray}
    \label{eq:cmats_weyl}
\end{align}
\vspace{2mm}

\noindent
Here for each matrix we show the row labels on the left \brk{which may start
with negative integers when $\muL < 0$}, and overall number of
rows and columns on the right and at the bottom respectively. We also highlight
with grey the row corresponding to the $\beta^{-1}$ term of the Laurent
expansion as it plays an important role for the global residue used in following.

\subsubsection{Companion tensor representation}
\label{sssec:companion_tensor}
We are now in position to combine the two matrix representations introduced
above into a single tensor representation that is going to be used for
intersection number computation later.
For simplicity, we will focus on the univariate case~\eqref{eq:ansatz_univariate},
as the generalization to multivariate~\eqref{eq:ansatz_multivariate} is
straightforward.

\paragraph{Step 1: Polynomial reduction}
The key idea is to rewrite the ansatz~\eqref{eq:ansatz_univariate} for an element of the
quotient ring~\eqref{eq:space_quotient} in a vector form
\begin{align}
    \psiL
    =
    \sum_{a = 0}^{\kappa - 1} \>
    z^a \> \psiL_a\brk{\beta}
    \equiv
    \begin{bNiceMatrix}[margin]
        1 & z & \Ldots[color = gr] &
        z^{\kappa - 1}
    \end{bNiceMatrix}
    \cdot
    \begin{bNiceMatrix}[margin]
        \psiL_{0}\brk{\beta}
        \\
        \psiL_{1}\brk{\beta}
        \\
        \Vdots[color = gr]
        \\
        \psiL_{\kappa - 1}\brk{\beta}
    \end{bNiceMatrix}
    \>,
    \label{eq:ctensor_ansatz_z}
\end{align}
where we regard the row vector of irreducible monomials $z^a$ on the left as a
basis \brk{of $\cQ$ viewed as a vector bundle} and collect the coefficients
$\psiL_a\brk{\beta}$ on the right.

Inside of the quotient ring~\eqref{eq:space_quotient}, multiplication by some polynomial
\begin{align}
    p\brk{z, \beta}
    =
    \sum_{a}
    \>
    z^a
    \>
    p_{a}\brk{\beta}
    \quad
    \wavy{1}
    \quad
    Q_p = \sum_a \> \brk{Q_{z}}^a \> p_a\brk{\beta}
    \label{eq:p_ctensor_z}
\end{align}
is captured by the companion matrix $Q_p\>$, constructed using the
basic monomial matrix~\eqref{eq:cmats_fin} \brk{see~\appref{app:code} for
an alternative method}.
Similarly, multiplication by some
rational function consisting of polynomial numerator $n\brk{z, \beta}$ and
denominator $d\brk{z, \beta}$ is expressed by
\begin{align}
    f\brk{z, \beta}
    =
    \frac{n\brk{z, \beta}}{d\brk{z, \beta}}
    \quad
    \wavy{1}
    \quad
    Q_f = Q_n \cdot \brk{Q_d}^{-1}
    \label{eq:ctensor_f_z}
    \>,
\end{align}
i.e. a product of the numerator and the inverse of the denominator
matrices \footnote{
    The order of operations here does not matter, as companion matrices $Q_f$
    for multiplication operators \brk{i.e. excluding $Q_{\partial_z}$} form a
    commutative algebra.
}.
Companion matrix $Q_f$ encodes the result of the polynomial reduction of the
product of the ansatz~\eqref{eq:ctensor_ansatz_z} and the rational
function~\eqref{eq:ctensor_f_z} as matrix-vector multiplication
\begin{align}
    \Bigfloor{f \> \psiL}_{\vev{\Baikov}}
    =
    \sum_{a_1 , a_2 = 0}^{\kappa - 1}
    z^{a_1}
    \>
    \bigbrk{Q_f}_{a_1 a_2}
    \>
    \psiL_{a_2}\brk{\beta}
    \equiv
    \begin{bNiceMatrix}[margin]
        1 & z & \Ldots[color = gr] & z^{\kappa - 1}
    \end{bNiceMatrix}
    \cdot
    Q_f
    \cdot
    \begin{bNiceMatrix}
        \psiL_0\brk{\beta}
        \\
        \psiL_1\brk{\beta}
        \\
        \Vdots[color = gr]
        \\
        \psiL_{\kappa - 1}\brk{\beta}
    \end{bNiceMatrix}
    \>.
\end{align}
Upon this substitution, the companion matrix representation of eq.~\eqref{eq:system_modular_univariate} reads:
\begin{align}
    Q_{\widehat{\nabla}_{-\omega}}
    \cdot \psi - \hatPhiR
    & = 0
    \>,
    \label{eq:deq_modular_univariate_Q}
\end{align}
where
\begin{align}
    Q_{\widehat{\nabla}_{-\omega}}
    \equiv
    {
        Q_{\partial_z \Baikov}\>\partial_\beta
        - Q_{\widehat{\omega}}
        + Q_{\partial_z}
    } \>.
    \label{eq:ctensor_nabla_univariate_Q}
\end{align}

\paragraph{Step 2: Series expansion}
Now let us include the Laurent series expansion in $\beta \to 0$ of the
ansatz~\eqref{eq:ctensor_ansatz_z} encapsulated in an infinite matrix of
coefficients
\begin{align}
    \psiL \Big|_{\beta \to 0}
    &=
    \sum_{a = 0}^{\kappa - 1} \>
    \sum_{n} \>
    z^a
    \>
    \beta^n
    \>
    \psiL_{a n}
    \\
    &\equiv
    \begin{bNiceMatrix}[margin]
        1 & z & \Ldots[color = gr] & z^{\kappa - 1}
    \end{bNiceMatrix}
    \cdot
    \Bigbrk{
        \ldots
        +
        \begin{bNiceMatrix}[margin]
            \psiL_{0 n}
            \\
            \psiL_{1 n}
            \\
            \Vdots[color = gr]
            \\
            \psiL_{\kappa - 1 \, n}
        \end{bNiceMatrix}
        \>
        \beta^n
        +
        \begin{bNiceMatrix}[margin]
            \psiL_{0 \, n + 1}
            \\
            \psiL_{1 \, n + 1}
            \\
            \Vdots[color = gr]
            \\
            \psiL_{\kappa - 1 \, n + 1}
        \end{bNiceMatrix}
        \>
        \beta^{n + 1}
        +
        \ldots
    }
    \\
    &\equiv
    \begin{bNiceMatrix}[margin]
        1 & z & \Ldots[color = gr] &
        z^{\kappa - 1}
    \end{bNiceMatrix}
    \cdot
    \begin{bNiceMatrix}[margin]
        \Ldots[color = gr] & \psiL_{0 n} & \psiL_{0 \, n + 1} & \Ldots[color = gr]
        \\
        \Ldots[color = gr] & \psiL_{1 n} & \psiL_{1 \, n + 1} & \Ldots[color = gr]
        \\
        & \Vdots[color = gr] & \Vdots[color = gr]
        \\
        \Ldots[color = gr] & \psiL_{\kappa - 1 \, n} & \psiL_{\kappa - 1 \, n + 1} & \Ldots[color = gr]
    \end{bNiceMatrix}
    \cdot
    \begin{bNiceMatrix}[margin]
        \Vdots[color = gr] \\ \beta^{n} \\ \beta^{n + 1} \\ \Vdots[color = gr]
    \end{bNiceMatrix}
    \>.
    \label{eq:ctensor_ansatz_beta}
\end{align}
Almost all the basic operators get represented in terms of factorized tensors.
The exception is the $z$-monomial multiplication operator: to derive its
representation we first expand the matrix $Q_z$ shown in
eq.~\eqref{eq:cmats_fin} as a linear polynomial in $\beta$ with matrix
coefficients $Q_{z, 0}$ and $Q_{z, 1}$, namely
\begin{align}
    z
    \quad
    \wavy{1}
    \quad
    Q_z = Q_{z, 0} + Q_{z, 1} \> \beta
    \>,
\end{align}
and only then substitute $\beta$ with the corresponding matrix $L_{\beta}$ from
eq.~\eqref{eq:cmats_weyl_infinite} or~\eqref{eq:cmats_weyl}.
\paragraph{Step 3: Global residue and Tensor algebra}
The global residue~\eqref{eq:global_residue} acts as a covector on the space of
$\psi_{an}\>$, and in the basis of~\eqref{eq:ctensor_ansatz_beta}, it simply
extracts the coefficient of the $\psi_{\kappa - 1,\, -1}$ component. This
extraction can be carried out by multiplying with the row vector
$R \equiv E_{\kappa - 1} \otimes E_{-1}$, where $E_j$ is a row vector with
a single 1 in the $j^\text{th}$ position.
The intersection numbers can then be evaluated by applying the following list of
\textit{substitution rules}:
\begin{alignat}{6}
    z &  \quad \wavy{1} \quad &
      & \ctensor{z} &
    = & \> &
    \mId \> & \otimes &
         \> & Q_{z, 0} + L_{\beta} \otimes Q_{z, 1} \>, &
    \nonumber
    \\
    \partial_z & \quad \wavy{1} \quad &
               & \ctensor{\partial_z} &
             = & \> &
       \mId \> & \otimes &
            \> & Q_{\partial_z} \>, &
    \nonumber
    \\
    \beta & \quad \wavy{1} \quad &
          & \ctensor{\beta} &
        = & &
     L_{\beta} \> & \otimes &
                  & \mId \>, &
    \label{eq:replace_beta}
    \\
    \partial_{\beta}      & \quad \wavy{1} \quad &
                          & \ctensor{\partial_\beta} &
                        = & &
    L_{\partial_\beta} \> & \otimes &
                          & \mId \>, &
    \nonumber
    \\
    \Res_{\vev{\Baikov}} & \quad \wavy{1} \quad &
                         & R &
                       = & &
       E_{\kappa - 1} \> & \otimes &
                         & E_{-1} \>, &
    \nonumber
\end{alignat}
where, in the case of restricted representation~\eqref{eq:cmats_weyl}, the
covector $R$ is just the unit row vector with the element 1 in the $\abs{\mu} \, \kappa$
position (and all the other elements are vanishing)
\begin{align}
    R \defas
    \NiceMatrixOptions{code-for-last-row = \scriptstyle}
    \begin{bNiceMatrix}[last-row, margin]
        \gr{0} & \Cdots[color = gr] & \gr{0}
        & 1 &
        \gr{0} & \Cdots[color = gr] & \gr{0}
        \\
        & & & \overset{\uparrow}{\abs{\muL} \, \kappa} & & &
    \end{bNiceMatrix}
    \>,
    \label{eq:R_covector}
\end{align}
which effectively encodes the outcome of the global residue in
eqs.~\eqref{eq:global_residue_univariate, eq:global_residue}.
Multiplication by the series expansion of function~\eqref{eq:ctensor_f_z}
\begin{align}
    f\brk{z, \beta}
    \Big|_{\beta \to 0}
    =
    \sum_{a \, n}
    \>
    z^a \, \beta^n
    \>
    f_{a n}
    \quad
    \wavy{1}
    \quad
    \ctensor{f} = \sum_{a \, n} \> \brk{\ctensor{z}}^a \cdot \brk{\ctensor{\beta}}^n
    \>
    f_{an}
\end{align}
is then repackaged into a rank-4 companion tensor $\ctensor{f}$ as
\begin{align}
    \Bigfloor{f \> \psiL}_{\vev{\Baikov}}
    \Bigg|_{\beta \to 0}
    &=
    \sum_{a_1 ,\, a_2 = 0}^{\kappa - 1}
    \sum_{n_1 ,\, n_2}
    z^{a_1} \> \beta^{n_1}
    \>
    \bigbrk{\ctensor{f}}_{n_1 n_2 \, a_1 a_2}
    \>
    \psiL_{a_2 n_2}\brk{\beta}
    \>,
\end{align}
where the $\ctensor{f}$ tensor is defined as,
\begin{align}
    \ctensor{f} =
    \sum_{a \, n} \>
    \mId \otimes \bigbrk{Q_{z, 0} + L_{\beta} \otimes Q_{z, 1}}^a
    \cdot
    \bigbrk{L_{\beta} \otimes \mId}^n
    \>
    f_{an}
    \>.
    \label{eq:f_ctensor_beta}
\end{align}

The substitution rules collected in this subsection complete the
\textit{dictionary} of rules required for the tensor representation
of the systems of differential equations~\eqref{eq:system_modular_tensor},
which lie at the core of the evaluation of {\it intersection numbers by tensor
algebra} method discussed next.

\subsection{Companion tensors for intersection numbers}
\label{ssec:ctensors_interx}
Restoring the notation of~\secref{sec:review} and using the
replacements~\eqref{eq:replace_beta},
the {\it companion tensor representation} of the univariate
system~\eqref{eq:global_residue_univariate, eq:system_modular_univariate},
required in the evaluation of the intersection numbers for differential
1-forms, reads
\begin{align}
    \vev{\phiL \,|\, \phiR}
    +
    R \cdot \ctensor{\phiL} \cdot \psiL
    & = 0
    \>,
    \label{eq:res_modular_univariate}
    \\
    \ctensor{\widehat{\nabla}_{-\omega}}
    \cdot \psiL - \hatPhiR
    & = 0
    \>,
    \label{eq:deq_modular_univariate}
\end{align}
where
\begin{align}
    \ctensor{\widehat{\nabla}_{-\omega}}
    \equiv
    {
        \ctensor{\partial_z \Baikov} \cdot \ctensor{\partial_\beta}
        - \ctensor{\widehat{\omega}}
        + \ctensor{\partial_z}
    } \>.
    \label{eq:ctensor_nabla_univariate}
\end{align}

A simple application of generation and solution of such a system can be found
in~\appref{app:example}.

Similarly, the {\it companion tensor representation} of the multivariate
system~\eqref{eq:global_residue, eq:system_modular},
required in the evaluation of the intersection numbers for differential $m$-forms,
becomes
\begin{align}
    \vev{\phiL^{\brkbf{m}} \,|\, \phiR^{\brkbf{m}}}
    +
    R \cdot \ctensor{
            \vev{
                \phiL^{\brkbf{m}}\vert h\supbbf{m - 1}
            }
    } \cdot \psiL^{(m)}
    & = 0
    \label{eq:res_modular}
    \>,
    \\
    \ctensor{\widehat{\nabla}_{\Omega^\vee}}
    \cdot \psiL^{\brk{m}} - \hatPhiR^{\brk{m}}
    & = 0
    \>,
    \label{eq:deq_modular}
\end{align}
where
\begin{align}
    \ctensor{\widehat{\nabla}_{\Omega^\vee}}
    \equiv
    {
            \ctensor{\partial_z \Baikov} \cdot \ctensor{\partial_\beta}
            + \ctensor{\widehat{\Omega}^\vee}
            + \ctensor{\partial_z}
        } \ .
    \label{eq:ctensor_nabla}
\end{align}
The overall structure of the system~\eqref{eq:res_modular, eq:deq_modular} can
be consolidated into the following inhomogeneous matrix system
\begin{align}
    \begin{bNiceArray}{ccw{c}{3cm}c}[margin]
        1 & \Block{1-3}{
        R \cdot \ctensor{
                \vev{
                    \phiL^{\brkbf{m}}\vert h\supbbf{m - 1}
                }
            }} & &
        \\
        \mzero & \Block{3-3}{
            \ctensor{\widehat{\nabla}_{\Omega^\vee}}
        } & &
        \\
        \Vdots[color = gr] & & &
        \\
        \mzero & & &
        \CodeAfter
            \tikz \draw[
                line width=.4pt, gr,
            ]
                (2-|1) -- (2-|5)
            ;
            \tikz \draw[
                line width=.4pt, gr,
            ]
                (1-|2) -- (5-|2)
            ;
    \end{bNiceArray}
    \cdot
    \begin{bNiceMatrix}[margin]
        \vev{\phiL^{\brkbf{m}} \,|\, \phiR^{\brkbf{m}}}
        \\
        \Block{3-1}{\psiL^{(m)}}
        \\
        \\
        \\
        \CodeAfter
            \tikz \draw[
                line width=.4pt, gr,
            ]
                (2-|1) -- (2)
            ;
    \end{bNiceMatrix}
    =
    \begin{bNiceMatrix}[margin]
        \mzero
        \\
        \Block{3-1}{\hatPhiR^{\brk{m}}}
        \\
        \\
        \\
        \CodeAfter
            \tikz \draw[
                line width=.4pt, gr,
            ]
                (2-|1) -- (2)
            ;
    \end{bNiceMatrix}
    \label{eq:ctensor_system}
\end{align}
for the augmented column of unknowns that unites the intersection number and
the ansatz together.
This system has to be solved \textit{only} for the first unknown -- the
intersection number~\cite{Chestnov:2022okt} \brk{see also \cite{Gasparotto:2023cdl}}.
We notice, that in practice,
this formulation proves to be highly robust, enabling solutions even in cases
when the system is
ill-defined, such as when the connection matrix has a
resonant spectrum \brk{i.e., eigenvalues that differ by integers}.

The tensor systems of equations presented in this section constitute the first
major result of this communication. They offer a new perspective on
evaluation of intersection numbers of differential $n$-forms, applicable in both
physics and mathematics.
The key novel feature of the proposed algorithm is its reliance on a companion
tensor representation of differential operators that effectively replaces
arithmetic operations of polynomial division with matrix multiplication 
operating \textit{directly} on the remainders of this division.

Next we will show a non-trivial application of our formalism.

\section{Decomposition of two-loop five-point massless planar integrals}
\label{sec:pentabox}
\begin{figure}[H]
    \centering
    \includegraphics[width=0.9\textwidth]{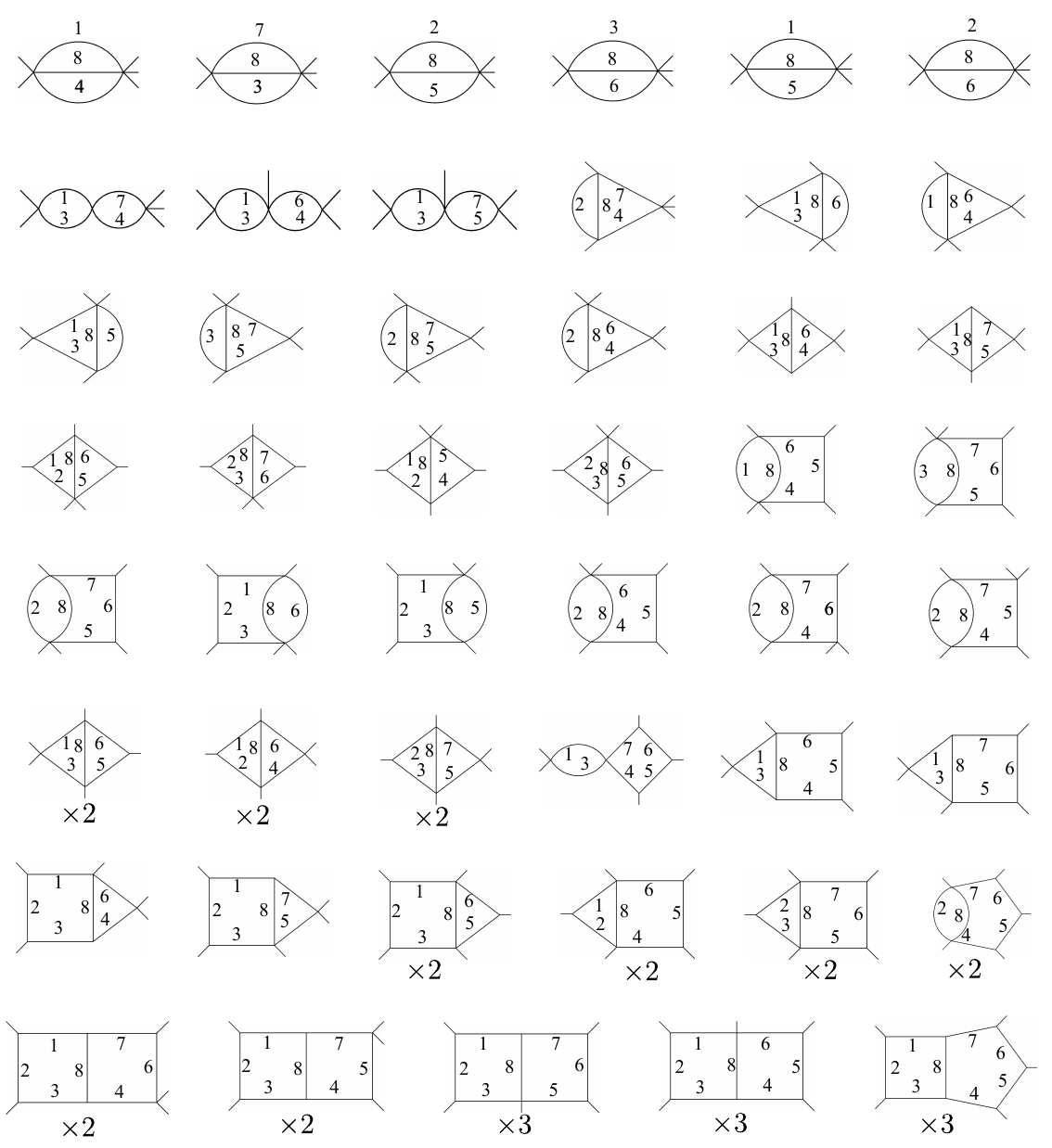}
\caption{
        The 47 sectors of the 62 master integrals defined in eq.~\protect\eqref{eq:mis}, corresponding to the massless two-loop five-points integral family (symmetry relations have not been applied).}
\label{fig:penta}
\end{figure}

The goal of this section, is to apply the algorithm described
in~\secref{sec:ctensor} for the numerical decomposition of massless two-loop five-point
functions in terms of master integrals.
A complete analytic decomposition is currently infeasible, as the most
computationally intensive parts of the problem require runtimes on the order of
a day or more. As such, pursuing a full analytic result is beyond the scope of
this work.
Further improvements in the efficiency of our implementation are left for
future investigation.
The integral family is defined in terms of 11 generalised denominators:
\begin{align}
z_1 &= k_1^2\>,
& z_2 &= (k_1{+}p_1)^2,
& z_3 &= (k_1{+}p_1{+}p_2)^2,
& z_4 &= (k_2{+}p_1{+}p_2)^2, \nonumber \\
z_5 &= (k_2{+}p_1{+}p_2{+}p_3)^2,
& z_6 &= (k_2{-}p_5)^2\>,
& z_7 &= k_2^2,
& z_8 &= (k_1{-}k_2)^2, \nonumber \\
 z_9 &= (k_2{+}p_1)^2,
& z_{10} &= (k_1{+}p_1{+}p_2{+}p_3)^2,
& z_{11} &= (k_1{-}p_5)^2,
\label{eq:penta_den}
\end{align}
as:
\begin{equation}
    I_{a_1a_2a_3 a_4 a_5 a_6 a_7 a_8 a_9 a_{10} a_{11}}  \ = \ \int \ \dd^{11}z\  u(\mathbf{z}) \frac{z_9^{-a_9}z_{10}^{-a_{10}}z_{11}^{-a_{11}}}{z_1^{a_1}z_2^{a_2}z_3^{a_3}z_4^{a_4}z_5^{a_5}z_6^{a_6}z_7^{a_7} z_8^{a_8}}
    \label{eq:penta}
\end{equation}
$z_9,z_{10},z_{11}$ are irreducible scalar products, and hence the set of relative boundaries is given by:
\begin{equation}
    D^\vee = V(\prod_{i=1}^8 z_i)\ .
\end{equation}
The kinematics is such that:
\begin{align}
    & p_i^2 \ = \ 0 \;, \ \
    s_{12} \ = \ (p_1{+}p_2)^2 \;, \ \
    s_{23} \ = \ (p_2{+}p_3)^2 \;, \nonumber \\
     & s_{34} \ = \ (p_3{+}p_4)^2 \;, \ \
     s_{45} \ = \ (p_4{+}p_5)^2 \;, \ \
       s_{51} \ = \ (p_5{+}p_1)^2 \;.
\end{align}
Therefore, our computation involves six external parameters
\begin{align}
    \vec{x} = \brc{s_{12}, s_{23}, s_{34}, s_{45}, s_{51}, d}
    \>.
\end{align}
We evaluate the intersection numbers numerically, restricting $\vec{x}$ 
to generic numerical values\footnote{
    The choice of kinematic region does not affect the performance of our
    implementation. We select generic numerical values that are away 
    from the singulartities of the process.
    For the proof-of-concept experiments, the integers appearing in the
    rational numbers are chosen to be relatively small to simplify
    the finite field reconstruction.
}, for example $\vec{x}_0 = \brc{1/3, -1/5, -1/7, 3/7, 5/13, 79/11}$.

This integral family has (before application of the symmetry relations) $\nu = 62$ master integrals, which we may pick\footnote{
    We may use any other basis of independent integrals as well.
} as depicted in \figref{fig:penta}, as:
\begin{align}
&J_{1}=I_{10010001000}, && J_{2}=I_{00100011000}, && J_{3}=I_{01001001000}, && J_{4}=I_{00100101000}, \nonumber\\
&J_{5}=I_{10001001000}, && J_{6}=I_{01000101000}, && J_{7}=I_{10110010000}, && J_{8}=I_{10110100000}, \nonumber\\
&J_{9}=I_{10101010000}, && J_{10}=I_{01010011000}, && J_{11}=I_{10100101000}, && J_{12}=I_{10010101000}, \nonumber\\
&J_{13}=I_{10101001000}, && J_{14}=I_{00101011000}, && J_{15}=I_{01001011000}, && J_{16}=I_{01010101000}, \nonumber\\
&J_{17}=I_{10110101000}, && J_{18}=I_{10101011000}, && J_{19}=I_{11001101000}, && J_{20}=I_{01100111000}, \nonumber\\
&J_{21}=I_{11011001000}, && J_{22}=I_{01101101000}, && J_{23}=I_{10011101000}, && J_{24}=I_{00101111000}, \nonumber\\
&J_{25}=I_{01001111000}, && J_{26}=I_{11100101000}, && J_{27}=I_{11101001000}, && J_{28}=I_{01011101000}, \nonumber\\
&J_{29}=I_{01010111000}, && J_{30}=I_{01011011000}, && J_{31}=I_{10101101000}, && J_{32}=I_{101011-11000}, \nonumber\\
&J_{33}=I_{11010101000}, && J_{34}=I_{110101-11000}, && J_{35}=I_{01101011000}, && J_{36}=I_{01101011-100}, \nonumber\\
&J_{37}=I_{10111110000}, && J_{38}=I_{10111101000}, && J_{39}=I_{10101111000}, && J_{40}=I_{11110101000}, \nonumber\\
&J_{41}=I_{11101011000}, && J_{42}=I_{11101101000}, && J_{43}=I_{111011-11000}, && J_{44}=I_{11011101000}, \nonumber\\
&J_{45}=I_{11011101-100}, && J_{46}=I_{01101111000}, && J_{47}=I_{01101111-100}, && J_{48}=I_{01011111000}, \nonumber\\
&J_{49}=I_{01011111-100}, && J_{50}=I_{11110111000}, && J_{51}=I_{11110111-100}, && J_{52}=I_{11111011000}, \nonumber\\
&J_{53}=I_{11111011-100}, && J_{54}=I_{11101111000}, && J_{55}=I_{111-11111000}, && J_{56}=I_{11101111-100}, \nonumber\\
&J_{57}=I_{11111101000}, && J_{58}=I_{111111-11000}, && J_{59}=I_{11111101-100}, && J_{60}=I_{11111111000}, \nonumber\\
&J_{61}=I_{11111111-100}, && J_{62}=I_{111111110-10} \ .
\label{eq:mis}
\end{align}
Our \soft{Mathematica} and \soft{FiniteFlow} implementation can take as input any integral of the type of eq.~\eqref{eq:penta}.
We identify two types of applications, to check and gauge the potential of our algorithm \begin{itemize}
    \item \textbf{Higher-rank integrals.}  We analyze the performance of our method when applied to the decomposition of high-rank integrals of the form:
    \begin{equation}
    I_n  \ = \ \int \ \dd^{11}z\  u(\mathbf{z}) \frac{z_9^{n}}{z_1z_2z_3z_4z_5z_6z_7 z_8}
    =
    I_{1 1 1 1 1 1 1 1 -n 0 0} \, ,
    \label{eq:penta_I}
\end{equation}
for the cases where the exponent $n$ takes the values $n=2,5,10,15,20$.
    \item \textbf{Higher-dimensions integrals.}  We investigate the decomposition of shifted-dimension integrals, generated by the insertion of integer powers of the Baikov polynomial in the numerator. 
    For the considered case, we decompose single and double power of $\mathcal{B}$,
        corresponding to a pentabox integral in $d + 2$ and $d + 4$ \cite{Tarasov:1996br, Lee:2009dh} dimensions, formed by a
        linear combination of 412 monomials up to rank 4, and 169744 monomials
        up to rank 8, respectively:
    \begin{equation}
    I_{\mathcal{B}^n} \ = \ \int \ \dd^{11}z\  u(\mathbf{z}) \frac{\mathcal{B}^n}{z_1z_2z_3z_4z_5z_6z_7 z_8}
        \, ,
        \quad \text{for $n = 1, 2$.}
    \label{eq:penta_baik}
\end{equation}
\end{itemize}
The results are expressed in terms of master integrals of eq.~\eqref{eq:mis} via a complete set of spanning cuts, as:
\begin{equation}
    I \ = \ \sum_{i=1}^{62} c_i \ J_i\>.
    \label{eq:pentabox_masters}
\end{equation}
The explicit expressions for the twist $u\brk{\mathbf{z}}$, the master
integrals $J_i$,
the bases of the (dual) co-homology groups,
and the numerical values of the coefficients $c_i$, computed by through the master decomposition formula \eqref{eq:masterdecomposition} in terms of intersection numbers, can be found in the supplementary material
file~\filename{pentabox\_massless.m}, and in the corresponding GitHub repository \href{https://github.com/GiacomoBrunello/pentabox_decomposition}{\faGithub}.\footnote{
    Computations were performed on a machine with an AMD EPYC 7282 16-Core
    Processor \brk{x86\_64 architecture}, featuring 32 logical CPUs at 2.79\,GHz and
    125\,GiB of system memory.
} 

The set of spanning cuts is given by the maximal cuts
of the first ten master integrals $\brc{J_1, \ldots, J_{10}}$ of eq.~\eqref{eq:mis}.
For each cut, after a choice of fibration has been chosen, the bases for each layer have been automatically generated using the algorithm described in~\secref{sec:FI_twist}.
 The cuts over $J_3, J_4$, and their symmetric partners $J_6, J_5$, represent the major bottleneck of this calculation.
 For $J_3$, choosing as fibration order $\{z_4,z_9,z_7,z_6,z_1,z_3,z_{10},z_{11}\}$, the dimensions
    of the internal bases read (from outer to inner): $\{31,28,12,4,2,2,1,1\}$.
 For $J_4$, choosing as fibration order
    $\{z_4,z_9,z_7,z_5,z_2,z_1,z_{10},z_{11}\}$, the dimensions of the internal
    bases read: $\{27,26,18,6,3,2,1,1\}$.
This complexity required the development of an advanced computational implementation in \soft{FiniteFlow}~\cite{Peraro:2016wsq,Peraro:2019svx} to construct companion matrices and perform tensor algebra. The details are described in~\appref{app:code}. 
The successful decomposition\footnote{
    The numerical decomposition coefficients have been successfully compared
    against those obtained via standard IBP techniques, employing
    \soft{NeatIBP}~\cite{Wu:2023upw} to generate systems and reducing them via
    \soft{FiniteFlow}, up to $n=10$.
} of a special monomial of rank $n=20$
and the complete decomposition
of the pentabox integrals in
$d+2$ and $d + 4$ dimensions stands as the second major result of this communication.
It completes and extends earlier studies exploring the decomposition of 2-to-2
and 2-to-3 2-loop integrals~\cite{Brunello:2023rpq,Frellesvig:2023rsk}, dealing
with the decomposition of an involved combination of integrals.
It marks a milestone in the decomposition of two-loop Feynman integrals into
master integrals through intersection numbers projection, indicating that this
novel method could become a valid option in presence of integrals with higher
numerators or denominator powers. 
Notably, our tests indicate that the complexity of intersection number
computations scales linearly
with the rank of the integrals' numerators {\brk{at least for the integrals shown in
eq.~\eqref{eq:penta_I}}}, therefore motivating further
applications and developments of intersection numbers-based methods to
decompose multi-scale integrals in higher-multiplicity cases 
within gauge theories or effective field theories, in tandem with the standard
IBP-based approaches. 

\newpage
\section{Conclusions}
\label{sec:outlooks}

Twisted period integrals appear in many applications of theoretical physics and
mathematics, making it pivotal to understand their nature. These
integrals belong to a finite-dimensional vector space whose geometry is
governed by an inner product, known as the intersection number.

In this work, we identified and explored the underlying tensor structures of this inner product,
empowering a novel computational procedure for evaluating
intersection numbers of multivariate differential forms.

Within the fibration-based approach, $n$-form intersection numbers are
calculated iteratively, one variable at a time, on separate layers.
The architecture of the cohomology group on each layer is shaped by the
connection and its geometric properties: the zeroes determine the dimension of
the cohomology group as a vector space, while the singularities characterize the
localization points of intersection numbers.
The connection also defines the differential equation for parallel
transport, with the intersection number arising from the sum of residues
of the local vector-valued holomorphic solutions
at the localization points.

The singularities of the connection can be
naturally collected in the vanishing set of a
suitably chosen interpolating polynomial ideal. A linear deformation of
the ideal generator with a new symbolic parameter then constitutes a
many-to-one change of variables. In this new coordinate system, intersection
numbers localize at a single point: the origin.

This perspective allowed us to introduce the ambient tensor space that organically
hosts the problem of evaluating intersection numbers.
The space consists of three components.
The family of local vector-valued solutions at localization points
is glued together and gauged to become an element of the quotient ring,
yielding the first two components of the tensor space.
Since only the local behavior of the solution around the origin
matters, it makes sense to work directly with the series expansions in
the new variable, forming the third factor of the tensor space.

Through this representation, the differential equation turns into a matrix operator
acting on the tensor space, while the residue becomes a simple covector, with
companion matrices serving as the building blocks.
As a result, intersection numbers can be computed using simple algebraic and
matrix operations, with finite field reconstruction significantly enhancing
performance.

Our novel companion-matrix-based framework has been applied to the numerical reduction of massless
two-loop five-point functions up to rank 20, requiring the computation of 11-form
intersection numbers. This result represents a milestone in the decomposition of
Feynman integrals and Euler-Mellin integrals into master integrals using intersection numbers.

\subsection*{Acknowledgements}
We acknowledge Gaia Fontana, Hjalte Frellesvig, Tiziano Peraro, and Andrzej Pokraka for comments on the manuscript.
We thank Giulio Crisanti, Hjalte Frellesvig, Federico Gasparotto, Manoj Kumar
Mandal, and Sid Smith for discussions and checks at various stages of the
project, including the two-loop five-point integral decomposition studies in
the context of our previous work \cite{Brunello:2023rpq,Frellesvig:2023rsk}.
G.B. thanks John Joseph Carrasco and the Northwestern University for
hospitality during the completion of this work.
V.C. is grateful to Gaia Fontana,
Saiei-Jaeyeong Matsubara-Heo,
Henrik Jessen Munch,
Tiziano Peraro, and Andrzej Pokraka for many insightful
discussions throughout the years.
P.M. wishes to acknowledge interesting discussions with Giulio Salvatori on global residues and companion matrices.
It is a pleasure to acknowledge the stimulating discussions among participants,
lecturers and organizers at the {\it Domoschool 2024 "Intersecting Feynman
Integrals"}, Domodossola, Italy.
G.B. research is supported by the European Research Council, under grant
ERC–AdG–88541, and by Università Italo-Francese, under grant Vinci.
The work of V.C. is supported by the European Research Council (ERC) under the
European Union’s Horizon Europe research and innovation programme grant
agreement 101040760 (ERC Starting Grant FFHiggsTop). 
We acknowledge the support of the INFN {\it Amplitudes} initiative.
We acknowledge CloudVeneto for the use of computing and storage facilities.
\newpage
\appendix

\addtocontents{toc}{\protect\setcounter{tocdepth}{1}} 

\section{Polynomial division as gauge transformation}
\label{app:polydiv}
Local solutions of~\eqref{eq:psiunivariate} are labeled by the points of
$V\brk{\Baikov}$. Let us assume for simplicity that there are no multiple poles, i.e.
\begin{align}
    V\brk{\Baikov} = \brc{
        p_{1}\brk{\beta}, \ldots, p_{\kappa}\brk{\beta}
    }
\end{align}
and all the points $p_{i}\brk{0}$ are different. The generating equation for
the ideal~\eqref{eq:poly_def}
\begin{align}
    \Baikov\brk{z} - \beta = 0
\end{align}
is a many-to-one polynomial map, such that $\Baikov\brk{p_i\brk{\beta}} \equiv
\beta$ for each $i$, see~\cite{Cacciatori:2021nli} for further discussion.

We may thus perform the coordinate change $z \mapsto \beta$ of the
connection~\eqref{eq:omega_def} near each point $p_{i}\brk{0}$ and organize the
local solutions to the differential equation~\eqref{eq:psiunivariate} into a
column vector
\begin{align}
    \psi =
    \begin{bNiceMatrix}[margin]
        \psiL_{p_{1}}\brk{\beta}
        &
        \psiL_{p_{2}}\brk{\beta}
        &
        \Ldots[color = gr]
        &
        \psiL_{p_{\kappa}}\brk{\beta}
    \end{bNiceMatrix}\tr
    \>,
\end{align}
which thus becomes a vector bundle of rank $\kappa$.
This vector of local solutions is related to the
ansatz~\eqref{eq:ctensor_ansatz_z} via the Vandermonde matrix
\brk{see also~\cite{Chestnov-Pokraka} for more applications}
\begin{align}
    \begin{bNiceMatrix}[margin]
        \psiL_{p_{1}}\brk{\beta}
        \\
        \psiL_{p_{2}}\brk{\beta}
        \\
        \Vdots[color = gr]
        \\
        \psiL_{p_{\kappa}}\brk{\beta}
    \end{bNiceMatrix}
    =
    \begin{bNiceMatrix}[margin]
        1 & p_{1} & p_{1}^2 & \Ldots[color = gr] & p_{1}^{\kappa - 1}
        \\
        1 & p_{2} & p_{2}^2 & \Ldots[color = gr] & p_{2}^{\kappa - 1}
        \\
        \Vdots[color = gr] & & & & \Vdots[color = gr]
        \\
        1 & p_{\kappa} & p_{\kappa}^2 & \Ldots[color = gr] & p_{\kappa}^{\kappa - 1}
    \end{bNiceMatrix}
    \cdot
    \begin{bNiceMatrix}[margin]
        \psiL_{0}\brk{\beta}
        \\
        \psiL_{1}\brk{\beta}
        \\
        \Vdots[color = gr]
        \\
        \psiL_{\kappa - 1}\brk{\beta}
    \end{bNiceMatrix}
    \>,
\end{align}
establishing a simple relation between the two descriptions via a gauge
transformation.

\section{A pedagogical example}
\label{app:example}
Consider a family of univariate twisted period integrals of the form
\begin{align}
    I_a &= \int_{\mathcal{C}} \! u \> \phiL,
    \qquad
    u = z^\rho \> \Baikov\brk{z}^\gamma,
    \qquad
    \varphi = \frac{\dd z}{z^{a}}
    \>,
\end{align}
where exponents $a \in \Integers$ and $\rho, \gamma \in \Complex$, and
$\Baikov\brk{z}$ the following quadratic monic polynomial:
\begin{align}
    \Baikov\brk{z} = b_0 + b_1 \, z + z^2
    \>,
    \label{eq:example_baikov}
\end{align}
with some generic coefficients $b_0, b_1 \in \Field$.
The connection~\eqref{eq:omega_def} is
\begin{align}
    \widehat{\omega}\brk{z} = \frac{\rho}{z} + \gamma \> \frac{\partial_z \Baikov}{\Baikov}
    \>,
    \label{eq:example_omega}
\end{align}
and its polar set is
\begin{align}
    \Poles_\omega = \brc{0} \cup V\brk{\Baikov}
    \>.
    \label{eq:example_poles}
\end{align}

To illustrate the computational scheme introduced in this communication, let us
focus on the intersection number:
\begin{align}
    \vev{\phiL \>|\> \phiR}
    =
    \vev{\dd z / \Baikov^2 \>|\> \dd z / \Baikov}
    \>.
    \label{eq:example_interx}
\end{align}
It can be easily seen, that the pole at $z = 0$ from the
set~\eqref{eq:example_poles} does not contribute to this intersection number,
so below we will only consider the vanishing locus $V\brk{\Baikov}$ for brevity.

\subsection{Step 1: Polynomial reduction}
The ideal generated by the polynomial~\eqref{eq:example_baikov} induces a
2-dimensional quotient space~\eqref{eq:space_quotient}
\begin{align}
    \cQ = \Span_{\Field\brk{\beta}}\bigbrk{{1, z}}
    \>.
\end{align}
This implies that the ansatz~\eqref{eq:ansatz_univariate} is a 2-dimensional
column vector
\begin{align}
    \psiL\brk{z, \beta}
    =
    \psiL_0\brk{\beta}
    +
    z \> \psiL_1\brk{\beta}
    \equiv
    \begin{bNiceMatrix}
        1
        &
        z
    \end{bNiceMatrix}
    \cdot
    \begin{bNiceMatrix}
        \psiL_0\brk{\beta}
        \\
        \psiL_{1}\brk{\beta}
    \end{bNiceMatrix}
    \label{eq:example_ansatz_z}
    \>,
\end{align}
and the companion matrices~\eqref{eq:cmats_fin} are explicitly
\begin{align}
    Q_{z}
    =
    \begin{bNiceArray}{cl}[margin]
        \mzero & - b_0 + \beta
        \\
        1 & -b_1
    \end{bNiceArray}
    \>,
    \quad
    Q_{\partial_z}
    =
    \begin{bNiceMatrix}[margin]
        \mzero & 1
        \\
        \mzero & \mzero
    \end{bNiceMatrix}
    \>.
    \label{eq:example_cmats}
\end{align}
The differential equation~\eqref{eq:system_modular_univariate} turns into a rank-2
linear system as in eq.~\eqref{eq:deq_modular_univariate_Q}
\begin{align}
 Q_{\widehat{\nabla}_{-\omega}} \cdot
    \begin{bNiceMatrix}
        \psiL_0\brk{\beta}
        \\
        \psiL_{1}\brk{\beta}
    \end{bNiceMatrix}
    -
    \begin{bNiceMatrix}
        \beta^{-1}
        \\
        \mzero
    \end{bNiceMatrix}
    = 0
    \>,
    \label{eq:example_system_z}
\end{align}
where:
\begin{align}
  Q_{\widehat{\nabla}_{-\omega}}
    \equiv
    {
        Q_{\partial_z \Baikov}\> \partial_\beta
        - Q_{\widehat{\omega}}
        + Q_{\partial_z}
    } \>.
\end{align}
Here the companion matrix representation of the $\widehat{\omega}$-term
follows directly from eq.~\eqref{eq:example_omega} using the method
of~\secref{sssec:companion_tensor}, giving
\begin{align}
   Q_{\hat{\omega}}
    = \frac{\gamma}{\beta} \> Q_{\partial_z \Baikov}
    + \rho \> Q_{1 / z}
   \>,
\end{align}
and the two appearing matrices are
\begin{align}
    Q_{\partial_z \Baikov}
    &\equiv b_1 \> \mId + 2 \> Q_z
    =
    \begin{bNiceMatrix}[margin]
        b_1 & 2 \brk{- b_0 + \beta}
        \\
        2 & -b_1
    \end{bNiceMatrix}
    \>,
    \\
    Q_{1 / z}
    &\equiv Q_{z}^{-1}
    =
    \frac{1}{- b_0 + \beta}
    \>
    \begin{bNiceMatrix}[margin]
        b_1 & - b_0 + \beta
        \\
        1 & \mzero
    \end{bNiceMatrix}
    \>.
\end{align}
Finally, to simplify the subsequent formulae we find it convenient to multiply the
system~\eqref{eq:example_system_z} from the left with the inverse matrix $Q_{\partial_z \Baikov}^{-1}$ that takes the following form:
\begin{align}
    Q_{\partial_z \Baikov}^{-1}
    =
    \frac{1}{\Delta + 4 \beta}
    \begin{bNiceMatrix}[margin]
        b_1 & 2 \brk{- b_0 + \beta}
        \\
        2 & -b_1
    \end{bNiceMatrix}
    \>,
    \qquad
    \Delta
    \defas b_1^2 - 4 b_0
    \>,
\end{align}
where we recognize $\Delta = \operatorname{Disc}_z\bigbrk{\Baikov\brk{z}}$
as the discriminant of the polynomial~\eqref{eq:example_baikov}.
The system thus becomes
\begin{align}
    \Bigbrk{
        \partial_\beta
        -
        \frac{\gamma}{\beta}
        -
        \rho \>
        Q_{\partial_z \Baikov}^{-1} \cdot Q_{1 / z}
        +
        Q_{\partial_z \Baikov}^{-1} \cdot Q_{\partial_z}
    } \cdot
    \begin{bNiceMatrix}
        \psiL_0\brk{\beta}
        \\
        \psiL_{1}\brk{\beta}
    \end{bNiceMatrix}
    -
    Q_{\partial_z \Baikov}^{-1}
    \cdot
    \begin{bNiceMatrix}
        \beta^{-1}
        \\
        \mzero
    \end{bNiceMatrix}
    = 0
    \>.
    \label{eq:example_system_z_standard}
\end{align}

To compute the intersection number~\eqref{eq:example_interx} we are only
interested in the $\beta \to 0$ series expansion of the
solution~\eqref{eq:example_ansatz_z} to the system~\eqref{eq:example_system_z},
which we turn to now.

\subsection{Step 2: Series expansion}
The general $\beta \to 0$ series expansion of the ansatz~\eqref{eq:ctensor_ansatz_beta}
specifies in the case of eq.~\eqref{eq:example_ansatz_z} to
\begin{align}
    \psiL\brk{z, \beta}
    \Big|_{\beta \to 0}
    =
    \sum_{a = 0}^1
    \sum_{n \ge 0}
    z^a
    \>
    \psiL_{a n}
    \>
    \beta^n
    \equiv
    \begin{bNiceMatrix}[margin]
        1
        &
        z
    \end{bNiceMatrix}
    \cdot
    \begin{bNiceMatrix}[margin]
        \Ldots[color = gr] & \psiL_{0 \, -1} &
        \psiL_{0 0} & \psiL_{0 1} & \Ldots[color = gr]
        \\
        \Ldots[color = gr] & \psiL_{1 \, -1} &
        \psiL_{1 0} & \psiL_{1 1} & \Ldots[color = gr]
    \end{bNiceMatrix}
    \cdot
    \begin{bNiceMatrix}[margin]
        \Vdots[color = gr] \\ \beta^{-1} \\ 1 \\ \beta \\ \Vdots[color = gr]
    \end{bNiceMatrix}
    \label{eq:example_ansatz_beta}
    \>.
\end{align}
For the specific case of the intersection number~\eqref{eq:example_interx}, the
values of the exponents~\eqref{eq:omega_expansion} are $\mu = -2$ and $\mu^\vee
= -1$, so it is enough to bound the ansatz above to the interval
$\brc{\beta^{-1}, 1, \beta}$. It can be easily seen, that the $\psi_{a\,, -1}$
are going to be fixed to 0, so we will drop them in the following for
simplicity.
The series expansion of the system~\eqref{eq:example_system_z_standard}
reads
\begin{tikzpicture}[overlay, remember picture]
    \fill[
        rounded corners,
        fill = red, fill opacity = .4
    ]
        ($(pic cs:mark1) + (-2pt, -12pt)$) rectangle ++(38pt, 30pt)
    ;
    \fill[
        rounded corners,
        fill = red, fill opacity = .2
    ]
        ($(pic cs:mark2) + (-2pt, -22pt)$) rectangle ++(146pt, 48pt)
    ;
    \fill[
        rounded corners,
        fill = red, fill opacity = .05
    ]
        ($(pic cs:mark3) + (-2pt, -6pt)$) rectangle ++(18pt, 20pt)
    ;
    \fill[
        rounded corners,
        fill = purple, fill opacity = .4
    ]
        ($(pic cs:rhs1) + (-2pt, -12pt)$) rectangle ++(23pt, 30pt)
    ;
    \fill[
        rounded corners,
        fill = purple, fill opacity = .2
    ]
        ($(pic cs:rhs2) + (-2pt, -12pt)$) rectangle ++(20pt, 30pt)
    ;
    \fill[
        rounded corners,
        fill = purple, fill opacity = .05
    ]
        ($(pic cs:rhs3) + (-2pt, -6pt)$) rectangle ++(18pt, 20pt)
    ;
\end{tikzpicture}
\begin{align}
\renewcommand{\arraystretch}{1.25}
    \Bigbrk{
        \>
        \tikzmark{mark1}{
            \partial_\beta
            -
            \frac{\gamma}{\beta}
        }
        +
        \tikzmark{mark2}{
            \frac{-\rho}{\Delta}
        }
        \>
        \begin{bNiceMatrix}[margin]
            \tfrac{2 b_0 - b_1^2}{b_0} & b_1
            \\
            -\tfrac{b_1}{b_0} & 2
        \end{bNiceMatrix}
        +
        \frac{1}{\Delta}
        \>
        \begin{bNiceMatrix}[margin]
            \mzeroRedd & b_1
            \\
            \mzeroRedd & 2
        \end{bNiceMatrix}
        +
        \tikzmark{mark3}{
            \ldots
        }
        \>
    } \cdot
    \lrbrk{
        \begin{bNiceMatrix}
            \psiL_{00}
            \\
            \psiL_{10}
        \end{bNiceMatrix}
        +
        \begin{bNiceMatrix}
            \psiL_{01}
            \\
            \psiL_{11}
        \end{bNiceMatrix}
        \>
        \beta
    }&
    \nonumber
    \\[.5cm]
    -
    \lrbrk{
        \>
        \tikzmark{rhs1}{
            \frac{1}{\beta \, \Delta}
        }
        +
        \tikzmark{rhs2}{
            \frac{-4}{\Delta^2}
        }
        +
        \tikzmark{rhs3}{
            \ldots
        }
        \>
    }
    \>
    \begin{bNiceMatrix}[margin]
        b_1
        \\
        2
    \end{bNiceMatrix}
    = \> 0 \>, \qquad&
    \label{eq:example_system_z_standard_series}
\end{align}
where the opacity of the colors represent the degree in $\beta$ of the
collected terms.

\subsection{Step 3: Global residue and Tensor algebra}
Now we substitue the individual $\beta$-dependent operators with their
matrix representatives~\eqref{eq:cmats_weyl}, for example
\begin{align}
    \partial_{\beta}
    \quad \wavy{1} \quad
    L_{\partial_\beta} =
    \begin{bNiceMatrix}[margin]
        \mzero & 0 & \mzero
        \\
        \mzero & \mzero & 1
        \\
        \mzero & \mzero & \mzero
    \end{bNiceMatrix}
    \>,
    \qquad
    \text{and}
    \qquad
    \beta^{-1}
    \quad \wavy{1} \quad
    L_{\beta^{-1}} =
    \begin{bNiceMatrix}[margin]
        \mzero & 1 & \mzero
        \\
        \mzero & \mzero & 1
        \\
        \mzero & \mzero & \mzero
    \end{bNiceMatrix}
    \>,
\end{align}
leading to the tensor representation of the differential operator
in eq.~\eqref{eq:example_system_z_standard_series} of the form
\vspace{.1cm}
\begin{tikzpicture}[overlay, remember picture]
    \fill[
        rounded corners,
        fill = red, fill opacity = .4
    ]
        ($(pic cs:T1) + (-1pt, -28pt)$) rectangle ++(210pt, 62pt)
    ;
    \fill[
        rounded corners,
        fill = red, fill opacity = .2
    ]
        ($(pic cs:T2) + (-2pt, -28pt)$) rectangle ++(150pt, 62pt)
    ;
    \fill[
        rounded corners,
        fill = red, fill opacity = .05
    ]
        ($(pic cs:T3) + (-2pt, -6pt)$) rectangle ++(18pt, 20pt)
    ;
\end{tikzpicture}
\begin{align}
    \renewcommand{\arraystretch}{1.25}
    \tikzmark{T1}{}
    \begin{bNiceMatrix}[margin]
        \mzeroRed & 0 & \mzeroRed
        \\
        \mzeroRed & \mzeroRed & 1
        \\
        \mzeroRed & \mzeroRed & \mzeroRed
    \end{bNiceMatrix}
    \otimes
    \begin{bNiceMatrix}[margin]
        1 & \mzeroRed
        \\
        \mzeroRed & 1
    \end{bNiceMatrix}
    +
    \begin{bNiceMatrix}[margin]
        \mzeroRed & 1 & \mzeroRed
        \\
        \mzeroRed & \mzeroRed & 1
        \\
        \mzeroRed & \mzeroRed & \mzeroRed
    \end{bNiceMatrix}
    \otimes
    \begin{bNiceMatrix}[margin]
        -\gamma & \mzeroRed
        \\
        \mzeroRed & -\gamma
    \end{bNiceMatrix}
    +
    \tikzmark{T2}{}
    \begin{bNiceMatrix}[margin]
        1 & \mzeroRedd & \mzeroRedd
        \\
        \mzeroRedd & 1 & \mzeroRedd
        \\
        \mzeroRedd & \mzeroRedd & 1
    \end{bNiceMatrix}
    \otimes
    \begin{bNiceMatrix}[margin]
        \tfrac{-\rho}{\Delta} \tfrac{2 b_0 - b_1^2}{b_0} & \tfrac{1 - \rho}{\Delta} b_1
        \\
        \tfrac{\rho}{\Delta} \tfrac{b_1}{b_0} & \tfrac{2\brk{1 - \rho}}{\Delta}
    \end{bNiceMatrix}
    +
    \tikzmark{T3}{\ldots}
    \label{eq:example_system_T_standard}
\end{align}
The full system~\eqref{eq:example_system_z_standard_series} thus turns into
\begin{align}
    \renewcommand{\arraystretch}{1.3}
    \begin{bNiceMatrix}[left-margin = 5pt]
        \CodeBefore
            \rectanglecolor{red!40}{1-3}{2-4}
            \rectanglecolor{red!40}{3-5}{4-6}
            \rectanglecolor{red!20}{1-1}{2-2}
            \rectanglecolor{red!20}{3-3}{4-4}
            \rectanglecolor{red!20}{5-5}{6-6}
            \rectanglecolor{red!5}{3-1}{6-2}
            \rectanglecolor{red!5}{5-3}{6-4}
        \Body
        \ast & \ast & -\gamma & \mzeroRed & \mzero & \mzero
        \\
        \ast & \ast & \mzeroRed & -\gamma & \mzero & \mzero
        \\
        \ast & \ast & \tfrac{-\rho}{\Delta} \tfrac{2 b_0 - b_1^2}{b_0} & \tfrac{1 - \rho}{\Delta} b_1 & -\gamma + 1 & \mzeroRed
        \\
        \ast & \ast & \tfrac{\rho}{\Delta} \tfrac{b_1}{b_0} & \tfrac{2\brk{1 - \rho}}{\Delta} & \mzeroRed & -\gamma + 1
        \\
        \ast & \ast & \ast & \ast & \ast & \ast &
        \\
        \ast & \ast & \ast & \ast & \ast & \ast &
        \CodeAfter
            \tikz \draw[
                line width=.4pt, red,
            ]
                (3-|1) -- (3-|7)
                (5-|1) -- (5-|7)
                (1-|3) -- (7-|3)
                (1-|5) -- (7-|5)
            ;
    \end{bNiceMatrix}
    \cdot
    \begin{bNiceMatrix}[margin]
        \mzero
        \\
        \mzero
        \\
        \psiL_{0 0}
        \\
        \psiL_{1 0}
        \\
        \psiL_{0 1}
        \\
        \psiL_{1 1}
        \CodeAfter
            \tikz \draw[
                line width=.4pt, gr,
            ]
                (3-|1) -- (3-|2)
                (5-|1) -- (5-|2)
            ;
    \end{bNiceMatrix}
    -
    \begin{bNiceMatrix}[margin = 3pt]
        \CodeBefore
            \rectanglecolor{purple!40}{1-1}{2-1}
            \rectanglecolor{purple!20}{3-1}{4-1}
            \rectanglecolor{purple!5}{5-1}{6-1}
        \Body
        \tfrac{b_1}{\Delta}
        \\
        \tfrac{2}{\Delta}
        \\
        \tfrac{-4 b_1}{\Delta^2}
        \\
        \tfrac{-8}{\Delta^2}
        \\
        \ast
        \\
        \ast
        \CodeAfter
            \tikz \draw[
                line width=.4pt, purple,
            ]
                (3-|1) -- (3-|2)
                (5-|1) -- (5-|2)
            ;
    \end{bNiceMatrix}
    = 0
    \>,
    \label{eq:example_system_T_matrix}
\end{align}
we hid the irrelevant entries of the system under the $\ast$ signs.  Next, to
obtain the matrix form~\eqref{eq:ctensor_system}, we augment
eq.~\eqref{eq:example_system_T_matrix} with the additional
equation~\eqref{eq:res_modular}.  This equation represents the global residue
by a dot product with the row vector~\eqref{eq:R_covector}
\begin{tikzpicture}[overlay, remember picture]
    \fill[
        rounded corners,
        fill = gr, fill opacity = .4
    ]
        ($(pic cs:Res1) + (-1pt, -10pt)$) rectangle ++(31pt, 25pt)
    ;
    \fill[
        rounded corners,
        fill = gr, fill opacity = .2
    ]
        ($(pic cs:phi1) + (-1pt, -8pt)$) rectangle ++(10pt, 20pt)
    ;
    \fill[
        rounded corners,
        fill = gr, fill opacity = .4
    ]
        ($(pic cs:Res2) + (-1pt, -12pt)$) rectangle ++(68pt, 30pt)
    ;
    \fill[
        rounded corners,
        fill = gr, fill opacity = .2
    ]
        ($(pic cs:phi2) + (-1pt, -46pt)$) rectangle ++(68pt, 98pt)
    ;
    \fill[
        rounded corners,
        fill = gr, fill opacity = .4
    ]
        ($(pic cs:Res3) + (-1pt, -12pt)$) rectangle ++(68pt, 30pt)
    ;
\end{tikzpicture}
\begin{align}
    \vev{\phiL \>|\> \phiR}
    &= -\> \tikzmark{Res1}{}\Res_{\vev{\Baikov}}\bigbrk{
        \> \tikzmark{phi1}{}\phiL
        \> \psiL
    }
    \nonumber
    \\
    &= - \>
    \tikzmark{Res2}{}
    \begin{bNiceMatrix}[margin]
        \mzero & 1 & \mzero & \mzero & \mzero & \mzero
        \CodeAfter
            \tikz \draw[
                line width=.4pt, gr,
            ]
                (1-|3) -- (2-|3)
                (1-|5) -- (2-|5)
            ;
    \end{bNiceMatrix}
    \cdot
    \tikzmark{phi2}{}
    \begin{bNiceMatrix}
        \mzero & \mzero & \mzero & \mzero & 1 & \mzero
        \\
        \mzero & \mzero & \mzero & \mzero & \mzero & 1
        \\
        \mzero & \mzero & \mzero & \mzero & \mzero & \mzero
        \\
        \mzero & \mzero & \mzero & \mzero & \mzero & \mzero
        \\
        \mzero & \mzero & \mzero & \mzero & \mzero & \mzero
        \\
        \mzero & \mzero & \mzero & \mzero & \mzero & \mzero
        \CodeAfter
            \tikz \draw[
                line width=.4pt, gr,
            ]
                (3-|1) -- (3-|7)
                (5-|1) -- (5-|7)
                (1-|3) -- (7-|3)
                (1-|5) -- (7-|5)
            ;
    \end{bNiceMatrix}
    \begin{bNiceMatrix}[margin]
        \mzero
        \\
        \mzero
        \\
        \psiL_{0 0}
        \\
        \psiL_{1 0}
        \\
        \psiL_{0 1}
        \\
        \psiL_{1 1}
        \CodeAfter
            \tikz \draw[
                line width=.4pt, gr,
            ]
                (3-|1) -- (3-|2)
                (5-|1) -- (5-|2)
            ;
    \end{bNiceMatrix}
    = - \>
    \tikzmark{Res3}{}
    \begin{bNiceMatrix}[margin]
        \mzero & \mzero & \mzero & \mzero & \mzero & 1
        \CodeAfter
            \tikz \draw[
                line width=.4pt, gr,
            ]
                (1-|3) -- (2-|3)
                (1-|5) -- (2-|5)
            ;
    \end{bNiceMatrix}
    \cdot
    \begin{bNiceMatrix}[margin]
        \mzero
        \\
        \mzero
        \\
        \psiL_{0 0}
        \\
        \psiL_{1 0}
        \\
        \psiL_{0 1}
        \\
        \psiL_{1 1}
        \CodeAfter
            \tikz \draw[
                line width=.4pt, gr,
            ]
                (3-|1) -- (3-|2)
                (5-|1) -- (5-|2)
            ;
    \end{bNiceMatrix}
    \>.
\end{align}

The linear system that involves the non-zero elements of the
ansatz~\cite{Chestnov:2022okt,Gasparotto:2023cdl} is obtained by removing the
first column and the last row from eq.~\eqref{eq:example_system_T_matrix}
\begin{align}
    \renewcommand{\arraystretch}{1.3}
    \begin{bNiceMatrix}[margin]
        \CodeBefore
            \rectanglecolor{gr!40}{1-2}{1-5}
            \rectanglecolor{red!40}{2-2}{3-2}
            \rectanglecolor{red!40}{3-3}{3-3}
            \rectanglecolor{red!40}{4-3}{5-4}
            \rectanglecolor{red!40}{5-5}{5-5}
            \rectanglecolor{red!20}{4-2}{5-3}
        \Body
        1 & \mzero & \mzero & \mzero & 1
        \\
        \mzero & -\gamma & \mzero & \mzero & \mzero
        \\
        \mzero & \mzeroRed & -\gamma & \mzero & \mzero
        \\
        \mzero & \tfrac{-\rho}{\Delta} \tfrac{2 b_0 - b_1^2}{b_0} & \tfrac{1 - \rho}{\Delta} b_1 & -\gamma + 1 & \mzero
        \\
        \mzero & \tfrac{\rho}{\Delta} \tfrac{b_1}{b_0} & \tfrac{2\brk{1 - \rho}}{\Delta} & \mzeroRed & -\gamma + 1
        \CodeAfter
            \tikz \draw[
                line width=.4pt, gr,
            ]
                (2-|1) -- (2-|6)
            ;
            \tikz \draw[
                line width=.4pt, gr,
            ]
                (1-|2) -- (6-|2)
            ;
    \end{bNiceMatrix}
    \cdot
    \begin{bNiceMatrix}[margin]
        \vev{\phiL \>|\> \phiR}
        \\
        \psiL_{0 0}
        \\
        \psiL_{1 0}
        \\
        \psiL_{0 1}
        \\
        \psiL_{1 1}
        \CodeAfter
            \tikz \draw[
                line width=.4pt, gr,
            ]
                (2-|1) -- (2-|5)
            ;
    \end{bNiceMatrix}
    =
    \begin{bNiceMatrix}[margin]
        \CodeBefore
            \rectanglecolor{purple!40}{2-1}{3-1}
            \rectanglecolor{purple!20}{4-1}{5-1}
        \Body
        \mzero
        \\
        \tfrac{b_1}{\Delta}
        \\
        \tfrac{2}{\Delta}
        \\
        \tfrac{-4 b_1}{\Delta^2}
        \\
        \tfrac{-8}{\Delta^2}
        \CodeAfter
            \tikz \draw[
                line width=.4pt, purple,
            ]
                (2-|1) -- (2-|5)
            ;
    \end{bNiceMatrix}
    \>,
    \label{eq:example_system_final}
\end{align}
which we solve for the value of the intersection number $\vev{\phiL \>|\>
\phiR}$ only.
The final result reads
\begin{align}
    \vev{\phiL \>|\> \phiR}
    = -\psiL_{1 1}
    = \frac{8}{\Delta^2 \brk{\gamma - 1}}
    + \frac{-4 b_0 + \Delta \rho}{b_0 \Delta^2 \brk{\gamma - 1} \gamma}
    \>,
\end{align}
in agreement with~\cite{Crisanti:2024onv}. This
computation follows closely our computer implementation based on
\soft{FiniteFlow}~\cite{Peraro:2019svx}, whose general structure is shown in~\figref{fig:data_graph}
and some details are summarized in the following appendix.

\section{Implementation details}
\label{app:code}
In this appendix we collect the details of our computer implementation of the
framework presented in~\secref{sec:ctensor}.

\paragraph{Intersection matrices}
The systems shown in~\secref{ssec:ctensors_interx} are not limited to
computation of a single given intersection number $\vev{\phiL \,|\, \phiR}$.
Indeed, in our implementation we construct and solve the same \textit{single}
linear system for a collection of $\brc{\phiL_1, \ldots, \phiL_A}$ and
$\brc{\phiR_1, \ldots, \phiR_B}$ \brk{subscripts here temporarily denote
different cocycles in a list and not projection components} to determine the
corresponding $A \times B$ intersection matrix in one go.
In the context of the example discussed in~\appref{app:example} above, that
corresponds to the
{\tikz[baseline] \node[rounded corners, fill=purple!40, text=black, anchor=text] {column};}
on the RHS becoming a wider matrix with $B$
columns, and the top
{\tikz[baseline] \node[rounded corners, fill=gr!40, text=black, anchor=text] {row};}
on the LHS growing into a taller matrix with $A$ rows.

\paragraph{Choice of the polynomial ideal}
There are several methods to construct the polynomial~\eqref{eq:poly_def}.
Importantly it must contain \brk{a subset of} the singularities $\Poles_{\Omega^{(m)}}$
of the connection matrix $\Omega\supbrk{m}$ or its dual $\Omega^{\vee \,
\brk{m}}$ at a given layer $m$.
To satisfy the crucial condition~\eqref{eq:omega_expansion} we may choose the
$\Baikov\brk{z}$ as the least common multiple \brk{LCM} of all the denominators
appearing in the $\Omega^\vee$ connection matrix, as proposed in~\cite{Brunello:2023rpq}.
Alternatively, we may process separately each irreducible factor in the LCM
with its highest multiplicity, and add up the individual contributions in the
end.
In this way only simple poles in the $\beta$-variable may appear in the
connection matrix, which dramatically simplifies the construction of the tensor
representation~\eqref{eq:ctensor_nabla} in practice.

\paragraph{The infinity}
One possible prescription for computing the residue contribution
at $z = \infty$ is to take $\Baikov\brk{z} \defas
z^n$, where $n$ is the highest order pole of $\Omega^\vee$, and use exactly
the same solving procedure as before.
Alternatively, we may bring the $z = \infty$ to some finite position
$z_\infty$ via the M\"obius coordinate change
\begin{align}
    z \mapsto \frac{z}{z - z_\infty}
    \>.
\end{align}
However, in practice, we have not observed significant performance improvements
with this strategy, primarily due to the growth of polynomials in intermediate
expressions. Therefore, a careful consideration will be necessary in future
implementations of the M\"obius transformation approach.

\paragraph{Polynomial reduction via the Sylvester matrix}
To build the companion matrices presented in~\secref{sssec:cmat} we may
also just directly polynomial reduce numerator and denominator of individual
entries of $\Omega^\vee$, $\phiL^{\brk{m}}$, and $\phiR^{\brk{m}}$, as an
alternative to eq.~\eqref{eq:p_ctensor_z}. To do that we
build the augmented Sylvester\footnote{
    The same idea applies to multivariate polynomial ideals, in which case the
    corresponding matrix bears the name of F.S.~Macaulay~\cite{Macaulay:1916}.
} matrix~\cite{Sylvester:1840, Cox2015} of some polynomial $p\brk{z} = p_0 + z \, p_1 + \ldots + z^n \, p_n$ we wish to reduce modulo the chosen
polynomial~\eqref{eq:poly_def}

\begin{align}
    {S}_{p, \Baikov}
    =
    \begin{bNiceMatrix}[parallelize-diags=false, margin]
        \CodeBefore[create-cell-nodes]
            \tikz \fill[
                fill = gr, fill opacity = .1,
                rounded corners
            ]
                (1-|10) rectangle (5-|14)
            ;
        \Body
        1 & & &
        & p_n & p_{n - 1} & \Ldots[color = gr] & & p_1 & p_0 & & &
        \\
        & \Ddots[color = gr] & &
        & & \Ddots[color = gr] & & & & & \Ddots[color = gr] & &
        \\
        & & 1 &
        & & & p_n & p_{n - 1} & \Ldots[color = gr] & & p_1 & p_0 &
        \\
        & & & 1
        & & & & p_n & p_{n - 1} & \Ldots[color = gr] & & p_1 & p_0
        \\
        & & &
        & 1 & b_{\kappa - 1} & \Ldots[color = gr] & b_1 & b_0 & & & &
        \\
        & & &
        & & 1 & b_{\kappa - 1} & \Ldots[color = gr] & b_1 & b_0 & & &
        \\
        & & &
        & & & & & & & & &
        \\
        & & &
        & & & & 1 & b_{\kappa - 1} & \Ldots[color = gr] & b_1 & b_0 &
        \\
        & & &
        & & & & & 1 & b_{\kappa - 1} & \Ldots[color = gr] & b_1 & b_0
        \CodeAfter
            \tikz \draw[
                gr,
                line cap = round,
                line width = 1.06pt,
                dash pattern= on 0pt off .18cm
            ]
                ($(7)!-.25!(8)$) -- ($(7)!1.25!(8)$)
            ;
            \tikz \draw[
                gr,
                line cap = round,
                line width = 1.06pt,
                dash pattern= on 0pt off .18cm
            ]
                (7-|11) ++(-.3, 0) -- ++($(8) - (7)$)
            ;
            \tikz \draw[
                line width=.2pt, gr,
            ]
                (5-|1) -- (5-|14)
                (1-|5) -- (10-|5)
                (1-|10) -- (10-|10)
            ;
            \tikz \draw[<->, gr, shorten <> = .5em]
                (1-|14) ++(4mm, 0)
                -- ($(5-|14) + (4mm, 0)$)
                node [midway, fill=white] {\scriptsize$\kappa$}
            ;
            \tikz \draw[<->, gr, shorten <> = .5em]
                (5-|14) ++(4mm, 0)
                -- ($(10-|14) + (4mm, 0)$)
                node [midway, fill=white] {\scriptsize$n$}
            ;
            \tikz \draw[<->, gr, shorten <> = .5em]
                (1-|10) ++(0, 3mm)
                -- ($(1-|14) + (0, 3mm)$)
                node [midway, fill=white] {\scriptsize$\kappa$}
            ;
    \end{bNiceMatrix}
    \label{eq:sylvester}
\end{align}
In the top right corner we highlighted the $\kappa \times \kappa$ block, which,
upon row reduction, will contain the companion matrix $\cmatQ{p}$ up to
transposition and reversing the order of rows and columns:
\begin{align}
    \fun{RowReduce}{S_{p, \Baikov}}
    =
    \begin{bNiceMatrix}[margin]
        \CodeBefore[create-cell-nodes]
            \tikz \fill[
                fill = gr, fill opacity = .1,
                rounded corners
            ]
                (1-|3) rectangle (2-|4)
            ;
        \Body
        \mId & & \tilde{Q}_{p}
        \\
        & \mId & \ast
        \CodeAfter
            \tikz \draw[
                line width=.2pt, gr,
            ]
                (2-|1) -- (2-|4)
                (1-|2) -- (4-|2)
                (1-|3) -- (4-|3)
            ;
    \end{bNiceMatrix}
    \Longrightarrow
    \cmatQ{p} = \bigbrk{\tilde{Q}_p}\tr
    \code{\slash\slash} \, \code{Reverse} \, \code{\slash\slash} \, \code{Map[Reverse]}
    \>.
\end{align}
It is also possible to find the companion matrix representation of a rational
function with this technique, see~\cite{GiulioThesis} for details.

\begin{figure}
    \centering
    \includegraphicsbox{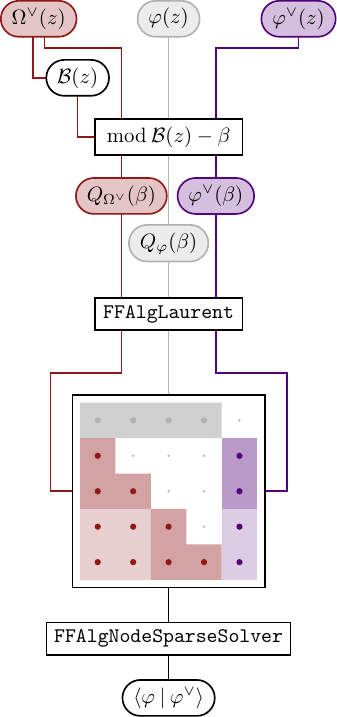}
    \caption{
        Flowchart of the data graph constructed in~\soft{FiniteFlow} for
        computing intersection numbers and described in~\appref{app:code}.
        Square in the middle represents the linear system of
        equations~\protect\eqref{eq:res_modular_univariate,
        eq:deq_modular_univariate} and~\protect\eqref{eq:res_modular,
        eq:deq_modular}, see eq.~\protect\eqref{eq:example_system_final} for an
        example.
    }
    \label{fig:data_graph}
\end{figure}

\paragraph{FiniteFlow data graph}
We find \soft{FiniteFlow}~\cite{Peraro:2016wsq,Peraro:2019svx}
to be an indispensable tool for our implementation
due to flexibility of its user interface and support for tensor constructions
that proved to be fundamental for the successful application
presented in~\secref{sec:pentabox}.
Indeed, many tensor operations like the outer product and contractions
can naturally be done within \soft{FiniteFlow}
via the nodes \code{FFAlgTakeAndAdd} and \code{FFAlgTakeAndAddBL} \brk{%
    with some automatization of the pattern
    creation~\href{https://github.com/vchestnov/utils}{\faGithub}%
}.
To compute intersection numbers using this technology, we convert the input data
$\brc{\Omega^\vee, \phiL, \phiR}$ first into companion matrix
representation~\eqref{eq:ctensor_f_z} utilizing the Sylvester matrix
technique~\eqref{eq:sylvester}, then series expand the produced matrices in
$\beta \to 0$ limit using the \code{FFAlgLaurent} node, to finally construct the
companion tensor representation~\eqref{eq:f_ctensor_beta}.
We then solve the system~\eqref{eq:system_modular_univariate, eq:system_modular}
with the \code{FFAlgNodeSparseSolver} node specifying only the
intersection number $\vev{\phiL \>|\> \phiR}$ itself as the needed unknown variable
to solve for.
We repeat the same procedure for the pole at infinity, add all the
contributions with the \code{FFAlgTakeAndAdd} node, and reconstruct only the final sum
of contributions to the given intersection number.
The schematic form of the data graph that we construct within
\soft{FiniteFlow} at each layer is shown in~\figref{fig:data_graph}.

\addtocontents{toc}{\protect\setcounter{tocdepth}{2}} 

\bibliographystyle{JHEP}
\bibliography{biblio}

\end{document}